%% file: document.tex
\RequirePackage{fix-cm}
\documentclass[smallextended]{svjour3}       
\smartqed  
\input{constants}

\begin{document}

\title{Generic Analysis of Model Product-Lines via 
Constraint Lifting}



\author{Andreas Bayha         \and
        Vincent Aravantinos 
}


\institute{Andreas Bayha \at
              fortiss GmbH, Research Institute of the Free State of Bavaria, Munich, Germany \\
              \email{bayha@fortiss.org}           
           \and
           Vincent Aravantinos \at
              CARIAD SE, Munich, Germany \\
              \email{vincent.aravantinos@cariad.technology}           
}

\date{}

\maketitle

\begin{abstract}
\input{text/abstract}
\end{abstract}

\section{Introduction}
\input{text/introduction}

\section{Running Example}
\input{text/example}

\section{Formal Basis: Definition of Non-Variable (Core) Models}
\input{text/formal-basis}
 
\input{text/dsml}

\subsection{Constraints}
\input{text/constraintLanguage}

\section{Conceptual Core: Model Product-Lines and Generic Product-Line Analysis by Constraint Lifting}
\input{text/variability-lifting}

\section{Exemplary Implementation Using SMT Solving}%
\input{text/smt-impl}

\section{Case Study: The \emph{SFIT} DSML for Manufacturing Planning}
\input{text/sfit}

\section{Evaluation}
\input{text/evaluation}

\section{Related Work}
\input{text/related-work}

\section{Conclusion and Future Work}
\input{text/summary}

\begin{acknowledgements}
We especially thank Lukas Hermanns and Raphael Gerdes from \bmw for the great cooperation during conception, implementation and modelling of \sfit and the engine models. 
They also made possible to present the screenshots and figures related to engine production in Section~\ref{sec:sfit}.   
\end{acknowledgements}


\section*{Conflict of interest}
The authors declare that they have no conflict of interest.

\bibliographystyle{spmpsci}      

\bibliography{document}

\end{document}

%% file: constants.tex
\usepackage{soul}

\usepackage[T1]{fontenc}
	
\usepackage[normalem]{ulem}
\usepackage{cancel}

\usepackage{xspace}

\usepackage{amssymb}
\usepackage{mathtools}
\usepackage{graphicx}

\usepackage{listings, lstautogobble}
 \usepackage{subcaption}
\usepackage{xcolor}

 \allowdisplaybreaks

\usepackage{ifthen}
\usepackage{amssymb} 
\usepackage{censor}
\usepackage{balance}

\usepackage{multirow}

\newboolean{reviewversion} 
\setboolean{reviewversion}{false}

\newboolean{showcomments} 
\setboolean{showcomments}{false} 
\ifthenelse{\boolean{showcomments}}
  {\newcommand{\nb}[2]{
    \fcolorbox{gray}{yellow}{\bfseries\sffamily\scriptsize#1}
    {$\blacktriangleright$#2$\blacktriangleleft$}
   }
    
  }
  {\newcommand{\nb}[2]{}
    
  } 
\newcommand\vincent[1]{\nb{Vincent}{\textcolor{blue}{#1}}}

\newcommand\andreas[1]{\nb{Andreas}{\textcolor{blue}{#1}}}

\newboolean{showtodo} 
\setboolean{showtodo}{true}
\ifthenelse{\boolean{showtodo}} 
  {\newcommand{\todo}[1]{
    \fcolorbox{gray}{yellow}{\bfseries\sffamily\scriptsize\textcolor{red}{TODO}}
    {$\blacktriangleright$\textcolor{red}{#1}$\blacktriangleleft$}
   }
  }
  {\newcommand{\todo}[1]{}}

\usepackage{comment}

\newboolean{showremoved} 
\setboolean{showremoved}{false} 
  
  \newboolean{showrewrite} 
\setboolean{showrewrite}{true} 

\newboolean{shownew} 
\setboolean{shownew}{false} 

  \ifthenelse{\boolean{showremoved}} 
  {
	  \newenvironment{remove}{\fcolorbox{gray}{red}{\bfseries\sffamily\scriptsize
	  remove}\color{red}$\blacktriangleright$}{$\blacktriangleleft$\ignorespacesafterend}
  }
  {
	  \usepackage{environ}
	\NewEnviron{hide}{}
	
	\ignorespacesafterend
  }
  
  \ifthenelse{\boolean{showrewrite}} 
  {\newenvironment{rewrite}{\fcolorbox{gray}{orange}{\bfseries\sffamily\scriptsize
  rewrite}\color{orange}$\blacktriangleright$}{$\blacktriangleleft$\ignorespacesafterend}}
  { \usepackage{environ}
	\NewEnviron{hideRew}{}
	
	\ignorespacesafterend}
  
  \definecolor{newcolor}{rgb}{0.0, 0.5, 0.0}
\ifthenelse{\boolean{shownew}} 
  {
	  \newenvironment{new}{\fcolorbox{gray}{newcolor}{\bfseries\sffamily\scriptsize
	  new}\color{newcolor}$\blacktriangleright$}{$\blacktriangleleft$\ignorespacesafterend}
	  }
  {
  \newenvironment{new}{}{\ignorespacesafterend}
  }

\lstnewenvironment{model}[1]
  {\noindent\textbf{Model}}
  {#1}

\newcommand*\eqdef{\stackrel{\text{def}}=}
\newcommand{\defeq}{\eqdef}

\newcommand{\newword}[2]{\newcommand{#1}{#2\xspace}}
\newword{\elemSet}{\mathbb{E}}
\newword{\objectSet}{\mathbb{O}}
\newword{\natSet}{\mathbb{N}}
\newword{\zSet}{\mathbb{Z}}
\newword{\multSet}{\mathbb{M}}
\newword{\boolSet}{\mathbb{B}}
\newword{\stringSet}{\mathbb{S}}
\newword{\typeSet}{\mathbb{T}}
\newword{\nameSet}{\mathbb{I}}
\newword{\classSet}{\mathbb{C}}
\newword{\attSet}{\mathbb{A}}
\newword{\assSet}{\mathbb{AS}}
\newword{\classFun}{\mathcal{C}}
\newword{\attFun}{\mathcal{AT}}
\newword{\assFun}{\mathcal{AS}}
\newword{\listSet}{\mathbb{L}}
\newword{\featuresSet}{F}
\newword{\mandatoryFeaturesSet}{\mathbb{F}_{\mathbb{M}}}
\def\fModel{\Phi}
\def\constr{\phi}
\newword{\cLang}{\mathcal{L}}

\newword{\lift}{\uparrow}
\newword{\bind}{\downarrow}
\newword{\config}{\lambda}
\newword{\confForm}{\config}
\newword{\id}{ID}
\newword{\idSet}{\mathbb {ID}}
\newword{\idMatch}{\stackrel{\idSet}{=}}
\newword{\parentFun}{parent}
\newword{\pcSet}{\mathbb{P}}
\newword{\geneva}{GeneVa}

\ifthenelse{\boolean{reviewversion}} 
{\newword{\sfit}{\censor{SFIT}}}
{\newword{\sfit}{SFIT}}
\ifthenelse{\boolean{reviewversion}} 
{\newword{\afocus}{\censor{AutoFOCUS3}}}
{\newword{\afocus}{AutoFOCUS3}}
\newword{\bmw}{the BMW Group}
\newword{\Bmw}{The BMW Group}
\newword{\wrt}{w.r.t.}
\newword{\true}{true}
\newword{\false}{false}
\newword{\requires}{\overset{req}{\longrightarrow}}
\newword{\requiresInst}{\overset{req}{\longmapsto}}
\newword{\excludes}{\overset{exc}{\longrightarrow}}
\newword{\excludesInst}{\overset{exc}{\longmapsto}}
\newword{\isOptional}{isOptional}

\newword{\bom}{BOM}
\newword{\none}{\perp}
\newword{\mL}{\mu L}
\newword{\microL}{$\mL$}
\newword{\tInt}{int}
\newword{\tBool}{bool}
\newword{\tString}{string}

\newword{\microSFIT}{$\mu SFIT$}

\lstdefinelanguage{microL}
{
  morekeywords={
    FunDecl,
    Fun,
    Function,
    function,
    VarDecl,
    Var,
    Program,
    INT,
    int,
    integer,
    float,
    bool,
    string
    STRING,
    BOOL,
    declare-datatypes
  },
  sensitive=false, 
  morecomment=[l]{//}, 
  morecomment=[l]{|}, 
  morecomment=[s]{[}{]}, 
  morecomment=[s]{/*}{*/}, 
  morestring=[b]" 
}

\definecolor{eclipseBlue}{RGB}{42,0.0,255}
\definecolor{eclipseGreen}{RGB}{63,127,95}
\definecolor{eclipsePurple}{RGB}{185,0,85}
 
\definecolor{darkblue}{rgb}{0,0,.6}
\definecolor{darkred}{rgb}{.6,0,0}
\definecolor{darkgreen}{rgb}{0,.6,0}
\definecolor{black}{rgb}{0,0,0}
\definecolor{red}{rgb}{.98,0,0}
\definecolor{blue}{rgb}{0,0,.98}
\definecolor{green}{rgb}{0,.98,0}
\definecolor{orange}{rgb}{1,.65,0}
\definecolor{hellgrau}{rgb}{0.90,0.90,0.90}
\definecolor{grau}{rgb}{0.50,0.50,0.50}

\lstdefinestyle{myListingStyle}{
   basicstyle=\scriptsize\ttfamily,
	 captionpos=b,
	 showspaces=false,
   showtabs=false,
   columns=fixed,
   frame=none,
	 numbers=left,
   numberstyle=\tiny\color{black},
   breaklines=true,
   showstringspaces=false,
   xleftmargin=5mm,
   tabsize=2,
   frame=single,
	 rulesepcolor=\color{black},
	 aboveskip=5mm,
	 commentstyle=\color{darkgreen},
	 identifierstyle=\color{black},
   keywordstyle=\bf\color{darkred},
   stringstyle=\color{blue},
	 extendedchars=true
 }

\lstdefinelanguage{smt}{
	sensitive=false,
  morekeywords={assert, forall, and, or, not, ite, exists,
  declare\-fun, declare\-const, define\-sort, declare\-datatypes, =>, =, <, >,
  bvugt, bvult, bvule, select, Int , bool, BitVec, seq.empty, seq.unit,
  seq.++, seq.contains, Seq, as}, alsoletter=-.+\=><, emph={:=},
	emphstyle=\bf\color{blue},
 	morecomment=[l]{;},
 	morecomment=[s]{'}{'},
	morestring=[b]",
}

\lstnewenvironment{smt}[2][mathescape=true]
	{\vspace{-.2cm}\lstloadlanguages{smt}\lstset{style=myListingStyle, language=smt, #1, #2}}
	{}

\makeatletter
\AtBeginDocument{%
  \let\c@figure\c@lstlisting
  \let\thefigure\thelstlisting
  \let\c@table\c@lstlisting
  \let\thetable\thelstlisting
  \let\ftype@lstlisting\ftype@figure 
  \let\ftype@table\ftype@figure 
}
\makeatother

\usepackage{tcolorbox}

\usepackage[framemethod=TikZ]{mdframed}
\mdfsetup{%
   middlelinewidth=.5pt,
   roundcorner=3pt}
   
\newenvironment{myexample}
{\vspace{.15cm}\begin{mdframed} \itshape \noindent \textsc{Example} for Definition \thedefinition. } 
{ \end{mdframed}\vspace{.15cm}}

\newenvironment{myexampleno}
{\vspace{.15cm}\begin{mdframed} \itshape \noindent \textsc{Example}. } 
{ \end{mdframed}\vspace{.15cm}}
 
 \makeatletter
\let\sv@thm\@thm
\def\@thm{\let\indent\relax\sv@thm}
\makeatother

%% file: text/abstract.tex
Engineering a product-line is more than just describing a product-line:
to be correct, every variant that can be generated must satisfy some constraints.
To ensure that all such variants will be correct (e.g. well-typed) there are only two ways:
either to check the variants of interest individually or to come up with a complex product-line analysis algorithm, specific to every constraint.

In this paper, we address a generalization of this problem:
we propose a mechanism that allows to check whether a constraint holds simultaneously for all variants which might be generated.
The main contribution of this paper is a function that assumes constraints that shall be fulfilled by all variants and generates (``lifts'') out of them constraints for the product-line. 
These lifted constraints can then be checked directly on a model product-line, thus simultaneously be verified for all variants. 
The lifting is formulated in a very general manner, which allows to make use of generic algorithms like SMT solving or theorem proving in a modular way.
We show how to verify lifted constraints using SMT solving by automatically translating model product-lines and constraints.
The applicability of the approach is demonstrated with an industrial case study, in which we apply our lifting to a domain specific modelling language for manufacturing planning.
Finally, a runtime analysis shows scalability by analyzing different model product-lines with production planning data from the BMW Group and Miele.

\vincent{State clearly that, even if it is not possible to define constraints
at product line level it is generally easy to get those at the variant
level.}

%% file: text/introduction.tex
\label{sec:dls}

\begin{figure}
 \centering
    \begin{subfigure}[b]{.65\columnwidth}
        \centering
        \includegraphics[width=\columnwidth]{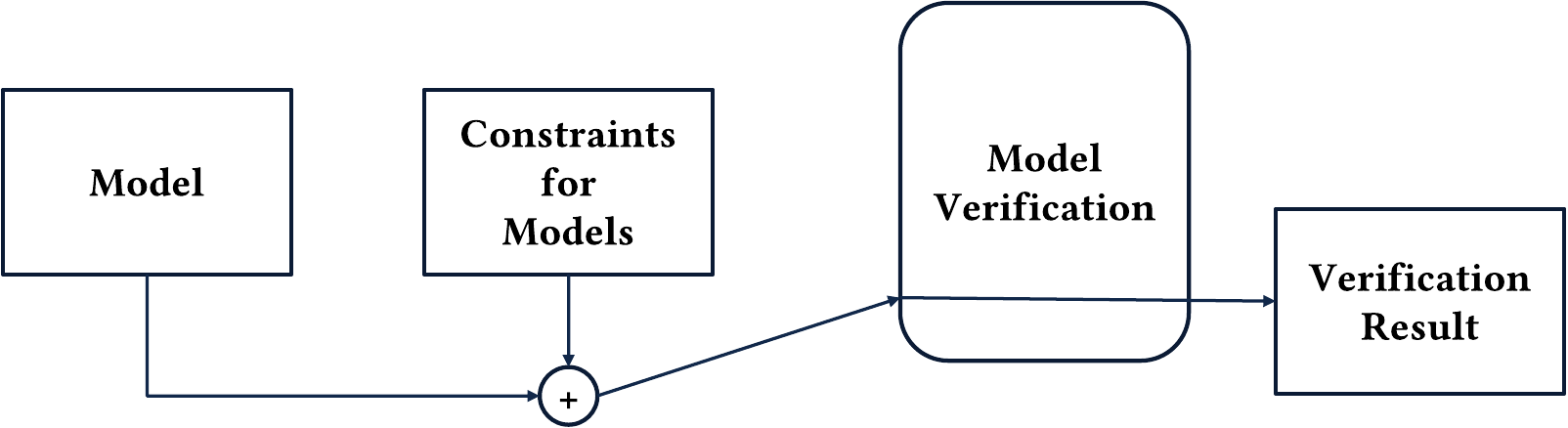}
        \caption{Approach for verifying a single (non-variable) model against a set of correctness constraints.}
        \label{fig:lifting_before}
    \end{subfigure}\\
    \vspace{.2cm}
    \begin{new}
    \begin{subfigure}[b]{0.7\columnwidth}
        \centering
        \includegraphics[width=\columnwidth]{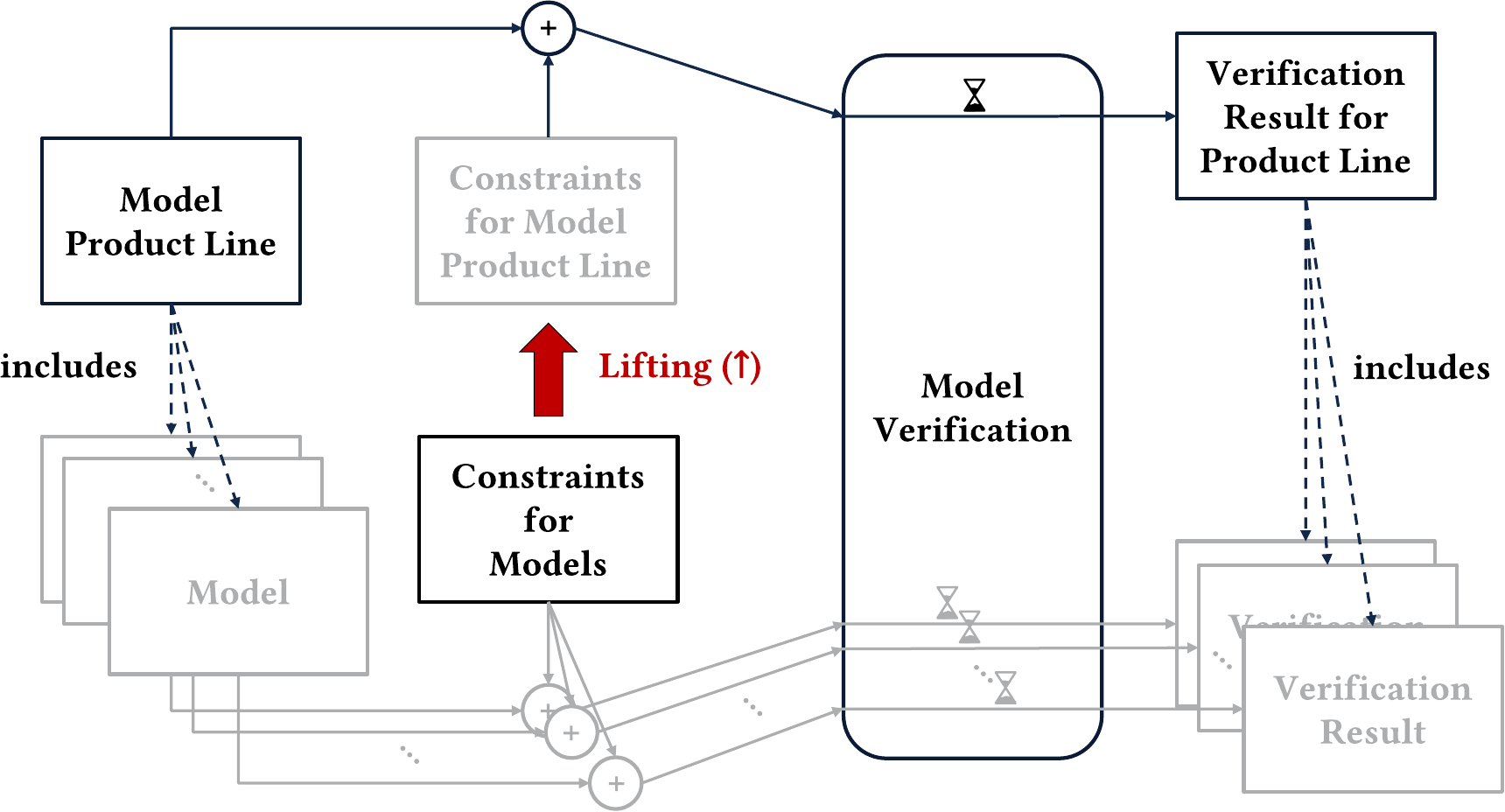}
        \caption{Our approach for model product-lines is to
       automatically lift constraints, such that they apply for the model
       product-line.
       With this we can verify the product-line by using the same verification without
        having to verify the single model variants.}
        \label{fig:lifting_after}
    \end{subfigure}
    \end{new}
\caption{Motivation and concept for our lifting approach.}
\label{fig:lifting}
\end{figure}

Many of today's products are produced as multiple different product variants. 
Some reasons for this variability in products arise from customers' demands, others from different regional situations within a global market. 
Also a company's individual portfolio management strategy is reflected, here.
To address this situation, (software) product-line engineering provides a methodology to develop multiple variants simultaneously, such that the same development artifacts can be reused among as many variants as required.

To do so, one classically develops a collection of reusable artifacts from which individual product variants can be generated as comfortably as possible -- ideally automatized. 
The development of such common artifacts (\textit{domain artifacts}) - the so-called \textit{domain engineering} - is for example explained by Pohl et al.~\cite{pohl2005software}.

The number of different variants that can be generated this way often gets very large, as it grows exponentially with the number of optional product features. 
Already 33 independent optional features allow $2^{33}$ (more than 8.5 billion) configurations -- an individual variant for every human being on Earth.
Though this demonstrates the potential power of this approach, it also causes problems, especially for verification. 
Classical testing requires to first generate the variants before checking them individually.

For many application domains however, one cannot afford to wait until a variant is generated to discover errors.
An example is a car for which the manufacturer only realizes during production that certain combinations of configuration options are incompatible.  
Furthermore, for the usually high number of variants, checking each and every single one individually would not even be possible.
The prominent example of the Linux Kernel comprises more than 10.000 features~\cite{lotufo2010evolution} -- compiling and testing every combination individually is impossible, here.
Also the study~\cite{rhein2018variability} by Rhein et al. illustrates this scalability issue with five real software product-lines. 
It demonstrates the ineffectiveness of such checks variant by variant, even for low numbers of variants.

Due to these problems of early verification and scalability, one would ideally like to check the domain artifacts of a product-line for correctness \emph{before} even generating variants. 
This approach is called \emph{product-line analysis} -- a technique and methodology that was mainly developed for software or software-intensive products.
This is why there are several analysis techniques that deal with the verification of specific software related issues -- type checking as by Kastner et al.~\cite{kastner2012type} is a prominent example, here. 

For model-based engineering however, developing such specific analysis algorithms is less efficient since different modeling languages might differ strongly from each other -- both in syntax and semantics.
While this makes it especially hard to transfer product-line analysis methods from one modeling language to another, the development of dedicated analysis algorithms for each modeling language is neither trivial nor efficient.
But especially such dedicated languages and home-grown tools have a big share of 38.2\% in the market of variability modelling tools in industrial practice, according to the survey of Berger et al.~\cite{berger2013survey}.
They all could potentially benefit from reusing such product-line analysis algorithms instead of implementing, debugging, and maintaining them again and again for every tool individually.

To overcome this problem with reusing \emph{specific} analysis mechanisms, there are \emph{generic} solutions in classical (i.e. \emph{non-variable}) model-based engineering. 
The usual solution hereby is the usage of generic \emph{constraint checkers} to verify \emph{correctness constraints} on a model, as illustrated in Figure~\ref{fig:lifting_before}. 
An example for such a constraint in systems engineering could be ``\textit{All incoming signals of all subsystems need to be provided values which match the type which is defined in the subsystems interfaces}''. 
An example for corresponding technologies could be the \textit{Object Constrain Language (OCL)} and corresponding OCL checkers.

If there however is variability in models, we deal with \emph{model product-lines} and several \emph{model variants}.
As indicated in Figure~\ref{fig:lifting_after}, the usage of generic constraint checkers for individual model variants becomes less efficient.
Here, every model variant will require time and effort to be verified individually and the same scalability and early verification issues arise, that are described above.

\newpage
\emph{In this work, we present a way to use generic verification mechanisms to implement a reusable product-line analysis for arbitrary modeling languages.}
To this end, we formally describe the conceptual core which is common to all model product-lines. 
We then define the consequences of variability in models in a procedure that we call \emph{symbolic binding}. 
Based on these considerations we define our approach that automatically adapts (``\emph{lifts}'') correctness constraints to take these consequences of variability into account. 
With this, it becomes sufficient to just specify correctness constraints for individual (i.e. non-variable) model variants.
The corresponding product-line analysis can then be obtained ``for free''. 
This can especially be helpful for the development of (new) domain specific modeling languages, but also existing languages can be extended with such analyses in this way.
 
Figure~\ref{fig:lifting_after} gives an overview about our approach.
As mentioned, the key idea is to automatically \emph{lift} constrains that are supposed to be correct for the individual model variants, such that they apply to the model product-line.
We do this \emph{constraint lifting} in such a way that a \emph{lifted} constraint holds for a model product-line iff the \emph{original} constraint will hold for all variants.
With this, a lifted constraint can be checked on the model product-line -- just as it would have been done for individual models.
Hereby classical generic verification mechanisms can be modularly utilized. 
Prominent possibilities are SMT solving or even theorem proving, if the base theories used in the constraint language are not decidable.
In case of SMT solvers, the verification result can also be a counterexample in form of a variant that violates a constraint.

After this formal definition of our approach, we present an exemplary implementation by usage of SMT solvers.
This implementation is then applied to an industrial case study with a modeling language for production planning. 
Therefore, we lift the language's constraints and translate them to SMT.   
Finally, the case study demonstrates the scalability with a runtime analysis on different model product-lines with production planning data from the BMW Group and the domestic appliances manufacturer Miele.

\begin{figure}
 \centering
        \includegraphics[width=\columnwidth]{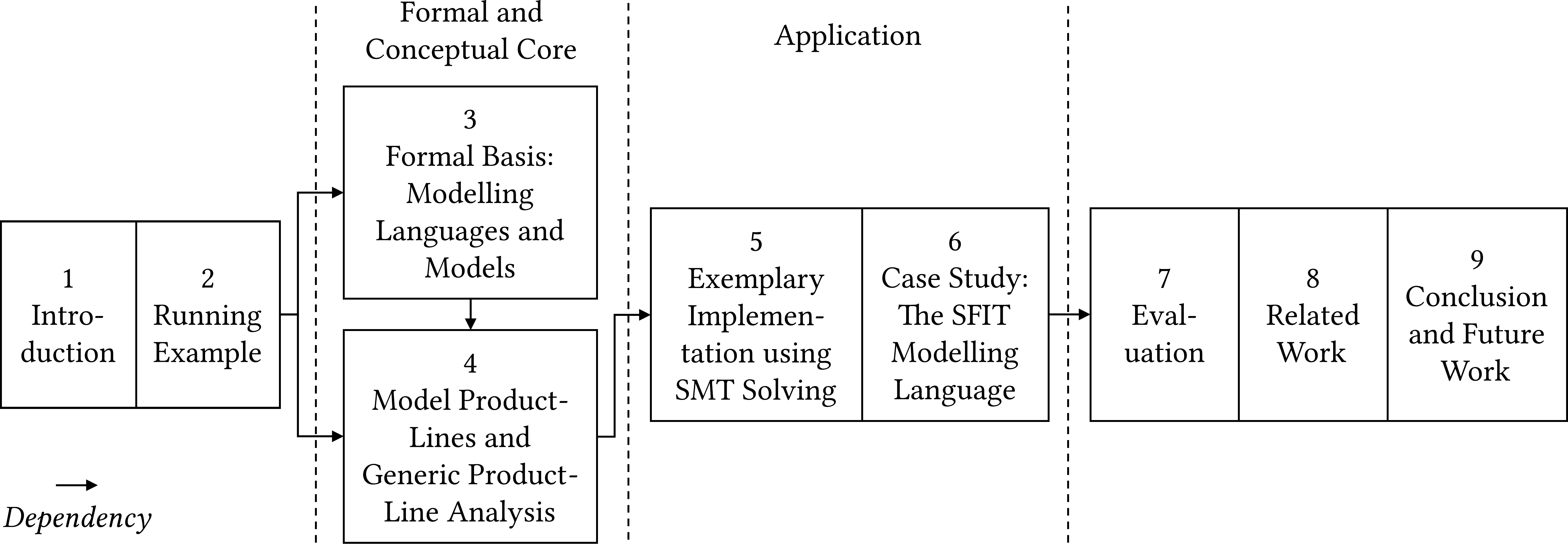}
        \caption{Overview about the structure of this paper and the dependencies between the chapters.}
\label{fig:outline}
\end{figure}

Figure~\ref{fig:outline} gives an overview about the organization of this paper:
 
First, we introduce a running example in Section~\ref{sec:examples}. 
This example is a small modelling language that will be used throughout the paper for illustrating purposes.

As the formal basis for defining our approach, we describe our notion of modeling and constraint languages in Section~\ref{sec:dsml}. 
This formalism will be required for the lifting and the translation to the verification mechanism.

The core of the paper is Section~\ref{sec:variability} in which we extend these notions by introducing variability in models. 
Here, we also present the \emph{symbolic binding} as a formalization of the effects of variability for models and the \emph{lifting function} for constraints which we consider our main contributions.

A possible implementation of our approach using SMT solving is presented in Section~\ref{sec:analysis}.

An industrial case study and an evaluation including a runtime analysis are presented in~Sections~\ref{sec:sfit} and~\ref{sec:evaluation}.

The paper concludes with the Sections~\ref{sec:related_work} on related work and ~\ref{sec:summary} on conclusion and future work.

%% file: text/example.tex
\label{sec:examples}

\begin{figure}
\centering
	\centering
	\includegraphics[width=.75\columnwidth]{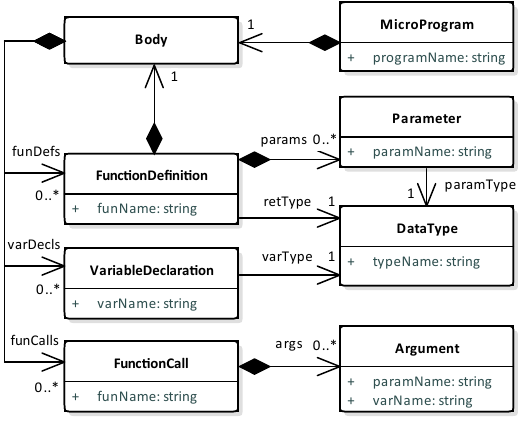}
	\caption{Metamodel for our running example: the modeling language \microL.}
	\label{fig:microLpp_metaModel}
\end{figure}

The approach for analyzing model product-lines in this paper is independent of a concrete modeling language. 
As a running example for the notions throughout the paper, we now introduce one modeling language for illustrating purposes.

It is simple language, that models function declarations, function calls and variables - we call this language \textit{Micro Language (\microL)}. 
The metamodel of \microL is given in Figure~\ref{fig:microLpp_metaModel}.

\begin{figure}
\centering
	\includegraphics[width=.7\columnwidth]{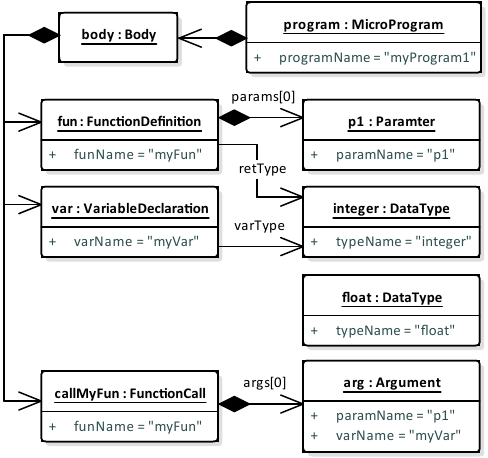}
	\caption{The example model ``myProgram1'' as an object diagram. The textual
	representation is given in Figure~\ref{lst:prog1}.}
\label{fig:microLpp}
\end{figure}

For a convenient presentation, we chose a textual syntax that is similar to C or Java. 
A small \microL program is given in Figure~\ref{lst:prog1} (textual) and~\ref{fig:microLpp} (object diagram).

In this language, a correctness constraint could be type correctness. 
This means, that the Arguments of all FunctionCalls need to have the same DataType, as the corresponding Parameter. 
$myProgram1$ of Figure~\ref{lst:prog1} is correct with respect to this constraint.
A counterexample is given in Figure~\ref{lst:prog2}. 
Here, the \textit{FunctionCall} will invoke an \textit{integer} function with a \textit{float} parameter in line 7 -- a typical typing error that a compiler would detect.  

Section~\ref{sec:constraint_language} will formally define correctness constraints for this example and Section~\ref{sec:variability_mod} will enhance the example with variability. 
Actually, there will also be Figure~\ref{lst:prog3} that shows a product-line, which contains the presented example models of Figures~\ref{lst:prog1} and~\ref{lst:prog2} as variants.

\begin{lstlisting}[float,caption={A simple example model in textual syntax. The corresponding object diagram can be found in Figure~\ref{fig:microLpp}. Some object names are annotated here in comments.},
label=lst:prog1, columns=flexible, numbers=left] 
Program myProgram1 			// program 
{ 											// body 
	Var integer myVar; 		// var
	
	myFun(								// callMyFun
		p1=myVar);  	 			// arg 
	
	Fun integer myFun(		// fun
			integer p1) 			// p1
		{ [...] }
}
\end{lstlisting}

\begin{lstlisting}[float,caption={A \microL program, that violates
the type correctness constraint for function calls.},label=lst:prog2, columns=flexible] 
Program myProgram2 {
	Var float myVar;
	
	// Here a float variable is passed to an integer
	// parameter. A violation of the type constraint 
	myFun(p1=myVar);
	
	Fun integer myFun(integer p1) {
		[...]	
	}
}
\end{lstlisting}

%% file: text/formal-basis.tex
\label{sec:dsml}

In this section, we introduce the necessary formal basis for the rest of the paper: 
metamodel, model, constraint language. 
These are the basic notions to deal with, when specifying a modeling language: 
constraints will be specified on the level of metamodels and are checked on models to verify their correctness.
This chapter does \emph{not} deal with variability:
here we first introduce the non-variable case as a baseline and later focus on variability and product-lines in a dedicated Section~\ref{sec:variability}.

%% file: text/dsml.tex
\subsection{Metamodels}
\label{sec:metamodel}

\def\MM{\mathbb{M}_{meta}}
\def\M{Inst}
\def\mLMeta{\mL}
\def\subMod{\sqsubseteq}
\def\sSubMod{\sqsubset}
We now formalize various necessary notions of metamodeling in an usual way. 
Let $\nameSet$ be a set of \emph{identifiers} (typically a set of strings in our examples). 
Let $\typeSet \subseteq \nameSet$ be a set of \emph{type identifiers} such that $\tInt, \tBool, \tString \in\typeSet$.
We call $\tInt, \tBool, \tString$ the \emph{basic types} of $\typeSet$ (we omit other potential basic types like floats for the sake of simplicity).
The domain for these basic types is $\boolSet$ for $\tBool$, $\zSet$ for $\tInt$ and the set of all strings $\stringSet$ for $\tString$. 
A \emph{multiplicity} is an element of $\multSet \defeq \{1, *\}$ (we omit other potential multiplicities for the sake of simplicity).
The set of \emph{attributes} $\attSet$ is defined as the set of tuples of $\typeSet\times\multSet$.
The set $\classSet$ of \emph{class bodies} is defined as the set of \emph{finite} functions of $\nameSet\to\attSet$.
The set $\MM$ of \emph{metamodels} is defined as the set of finite functions of $\typeSet\to\classSet$.
A metamodel $M \in \MM$ is \emph{well-defined} iff every type referred in $M$ is also a type defined in $M$.
We will only consider well-defined metamodels in the following.
Note that we formalize associations slightly different than often done, by formalizing them simply like the attributes above, which can have a non-basic type.
This makes the formalization simpler and does not entail any loss of generality.

\begin{myexampleno} 
	For the metamodel of Figure~\ref{fig:microLpp_metaModel} this means:
	\begin{align*}
		&int, string, Body, FunctionDefinition, funName &&\in \nameSet\\
		&int, string, Body, FunctionDefinition &&\in \typeSet\\
		&(string, 1), (Body, 1), (FunctionDefinition, *) &&\in \attSet\\
		&FunctionDefinition_{cb}, MicroProgram_{cb}&&\in \classSet\\
	\end{align*}	
	\vspace{-.7cm}
	\begin{align*}
		&FunctionDefinition_{cb} =\langle funName \mapsto (string,1),\\
		&\hspace{.7cm}params \mapsto (Parameter,*), retType \mapsto
		(DataType,1)\rangle\\
	&FunctionDefinition_{cb} (params)= (Parameter,*)\\
	&MicroProgram_{cb} = \langle programName \mapsto (String, 1),
	\ldots\rangle\\
	\end{align*}
	\vspace{-.7cm}	
	\begin{align*}
	&\mLMeta\in\MM\\
	&\mLMeta = \langle FunctionDefinition \mapsto FunctionDefinition_{cb},\\
		&\hspace{1cm}MicroProgram \mapsto MicroProgram_{cb}, \ldots\rangle\\
		&\mLMeta (FunctionDefinition)= FunctionDefinition_{cb}
	\end{align*}
\end{myexampleno}

\subsection{Core Models}
\label{sec:model}

We now define instances of a metamodel as well as the corresponding notions. 
In order to distinguish between non-variable \emph{models} and the \emph{model product-lines} of Section~\ref{sec:variability}, we denote \emph{models} without variability as \emph{core models}.

\begin{definition}[Instances]
Let $M \in \MM$ be a metamodel and $t\in\typeSet$ be a type defined in $M$.
The set $\M_M(t)$ of the \emph{instances} of $t$ in $M$ is inductively defined as follows:
	
\begin{align*}
\M_M(t) \defeq \begin{cases}
\boolSet &\text{if } t = bool\\
\zSet &\text{if } t = int\\
\stringSet &\text{if } t = string\\
\mathbb{O}_M(t) & \text{else}\\
\end{cases}\\
\end{align*}
 \end{definition}
 
 where the set $\mathbb{O}_M(t)$ of objects for a type $t$ in metamodel $M$ is defined as:
 
 \begin{align*}
 \mathbb{O}_M(t) \defeq \{ f | f(id) \in \M^{\attSet}_M(a) \text{ s.t. } &\\
\hspace{4mm}\exists id \in \nameSet, a \in \attSet, c \in \classSet.&\\
\hspace{4mm} M(t) = c \wedge c(id) = a\}
 \end{align*} 
 with $\M^{\attSet}_M(a)$ for an attribute $a = (t, m)$ defined as: 
 \begin{align*}
 \M^{\attSet}_M((t, m)) &\defeq 
 \begin{cases}
 \M_M(t) &\text{if } m = 1\\
 \mathbb L (\M_M(t))&\text{if } m = *\\
 \end{cases} 
\end{align*}
where $\mathbb L(S)$ denotes the set of lists of elements of a set.

We write $\M_M$ for the set of \emph{all}  models of a metamodel $M$, i.e. $\M_M \defeq \{m | \exists t \in \typeSet. m \in \M_M(t)\}$.
 
\begin{myexample}
The \microL model of Figure~\ref{fig:microLpp} would be: 
\begin{align*}
 &\M_{\mLMeta} (bool) = \{true,\ false\}\\
 &\langle typeName \mapsto\text{``float''}\rangle \in \M_{\mLMeta}
 (DataType)\\
 &\langle typeName \mapsto
 \text{``integer''} \rangle \in \M_{\mLMeta} (DataType)\\
 &\langle typeName \mapsto \text{``integer''} \rangle \in \M_{\mLMeta}\\
 &\langle funName \mapsto \text{``myFun''},\\
 &\ \ \ retType \mapsto \langle typeName \mapsto \text{``integer''}
 \rangle,\\
 &\ \ \ params \mapsto
 [\langle paramName \mapsto \text{``i1''},\\
 & \hspace{1.9cm}paramType
 \mapsto \langle typeName \mapsto \text{``integer''} \rangle
 \rangle]\\
 & \ \rangle \in \M_{\mLMeta} (FunctionDefinition)
\end{align*}
\end{myexample}

%% file: text/constraintLanguage.tex
\label{sec:constraint_language}

After we have introduced metamodels and their models, we continue with constraints. 
Constraints allow to restrict the set of instances of a metamodel which are considered valid.
For example, think back to the \microL program $myProgram2$ of Figure~\ref{lst:prog2} with the incorrectly typed function call:
even though it is an instance of the \microL metamodel, it is invalid due to the type error.

In metamodeling, constraints are classically expressed by OCL invariants.
Yet, not all modelling languages are based on UML.
A prominent example is the upcoming \emph{SysML v2} which will be based on the \textit{Kernel Modeling Language (KerML)}~\cite{kerml2021,sysmlv2spec}.
Hence, we want to stay independent of specific technologies and formalize a notion of constraints which is independent of any language.
At the same time, we also want to take care that our formalization is expressive enough to capture the means of such popular constraint languages as OCL.
Because of this, we decided to use first order (predicate) logic (FOL) for formalizing constraints, since it has been shown how OCL invariants can be translated to FOL.
At this point we refer to literature on such translations as Beckert et al. in~\cite{beckert2002translating} or Kuhlmann and Gogolla in~\cite{kuhlmann2012uml}.
With this, constraints could be formulated in OCL or any other language and then translated to FOL to be used with our analysis.   
As a pleasant side effect, this also simplifies the algorithms presented in this paper.

An essential aspect is that we parameterize our formalization with a \emph{base (first-order) theory} defining the atoms of the language. 
For instance, one can consider atoms that allow list expressions like $arguments.size = parameters.size$, or arithmetic expressions, etc.. 
We will also see these two theories in the SMT implementation of Section~\ref{sec:analysis}.
Yet, this base theory is considered domain-specific and therefore left undefined for our constraint formalization, which focuses only on \emph{composing} such atoms into complex constraints.

\def\lP{$<$}
\def\rP{$>$}
\newword{\gQUANT}{\text{{\lP}QEXPR\rP}}
\newword{\gDECL}{\text{{\lP}DECL\rP}}
\newword{\gSET}{\text{{\lP}SET\rP}}
\newword{\gVAR}{\text{{\lP}VAR\rP}}
\newword{\gEXPR}{\text{{\lP}EXPR\rP}}
\newword{\gQVAR}{\text{{\lP}NAV\rP}}
\newword{\gATOM}{\text{{\lP}ATOM\rP}}
\begin{definition}[Constraint Language $\cLang$]
\label{def:constrLang}
Let $B_1,\ldots,B_k$ be a set of base theories.
The constraint language $\cLang(B_1,\ldots,B_k)$ (or simply $\cLang$ when $B_1,\ldots,B_k$ are clear from the context) is defined by the following grammar:
{\small 
\begin{align*}
\cLang &\defeq &&\gQUANT\\
\gQUANT &\defeq &&\forall \gVAR \in \gSET :\gEXPR\\
\gVAR &\defeq &&[a-zA-Z]^{+}\\
\gSET &\defeq &&\typeSet \ |\ \gQVAR\\
\gQVAR &\defeq &&\gVAR\ |\ \gQVAR.\nameSet\\
	\gEXPR &\defeq &&\gATOM\ |\ \gQUANT\ |\ \neg\ \gEXPR\ |\\
& && \ \gEXPR \vee \gEXPR\\
  \gATOM &\defeq &&A_1 \ |\ ... \ |\ A_j
\end{align*}
}
where $A_1,...,A_j$ are arbitrary atoms of the base theories $B_1,\ldots,B_k$.
An example for such an atom, would again be the afore mentioned term $arguments.size = parameters.size$. 
\end{definition}

For brevity we only defined the $\forall$ quantifier and the Boolean operators $\vee$ and $\neg$ -- of course this is not a limitation and in the following, we also use $\exists,\ \wedge$ and $\implies$ as ``syntactic sugar''.

\begin{myexample}
\label{exa:constraints}
$\cLang$ constraints for \microL could be:

``All function names are unique'':
 \begin{align*} 
\forall &f_1, f_2 \in FunctionDefinition:\\
&((f_1.funName = f_2.funName) \implies
(f_1 = f_2)) 
\end{align*}
All arguments use only variables, which are defined:
 \begin{align*}
\forall &a \in Argument: \exists v \in VariableDeclaration:\nonumber\\
&a.varName = v.varName
\end{align*}
``All types of all variables used in all arguments of all calls, match with the
type of the respective parameters'':\pagebreak
\begin{align*}
\forall &F_{call} \in FunctionCall: \forall a \in F_{call}.args:\nonumber\\
\exists &F_{def} \in FunctionDefinition: \forall p \in
F_{def}.params:\nonumber\\
\nonumber \exists &v \in VariableDeclaration:\\
&a.paramName = p.paramName\ \wedge \nonumber\\
&a.varName = v.varName \implies \nonumber\\
&v.varType = p.paramType
\end{align*}
\end{myexample}

%% file: text/variability-lifting.tex
\label{sec:variability}

In the previous section, we described modeling languages with metamodels and constraints. 
Hereby (core) models are the instances of metamodels that shall fulfill all specified constraints.

Whenever there is a need for several \emph{variants} of a model, one usually does not want to maintain several copies of it, individually. 
Instead, one can follow the concept of \emph{domain engineering} and systematically capture the respective variability within one model. 
The variants can then be automatically generated from this model with variability, whenever needed. 
With such variability in a model, we speak of a model \emph{product-line}.

This section will first extend the notion of core models to model product-lines.  
After that, it will present our approach of how constraints can simultaneously be checked for all variants of such a model product-line.  

\subsection{Variability in Models}
\label{sec:variability_mod}

To formalize variability, we utilize the usual feature model notion as introduced by Kang et al.~\cite{kang1990feature}.
Let $\featuresSet$ be the set of all features, then $\fModel_{\featuresSet}$ a feature model over $\featuresSet$. 
As there is plenty of work on how to formalize feature models, we do not further go into details and consider $\fModel_{\featuresSet}$ as being a propositional logic formula that encodes which feature configurations are allowed.
For more details on how to formalize $\fModel_\featuresSet$ itself and on how to translate feature models to propositional logic formulas, see e.g. Batory~et~al.~\cite{batory2005feature} or Benavides~et~al.~\cite{benavides2007fama}. 
Section~\ref{sec:sfit} will also show an example for such a formula.

\begin{figure}
\centering
\includegraphics[width=.75\columnwidth]{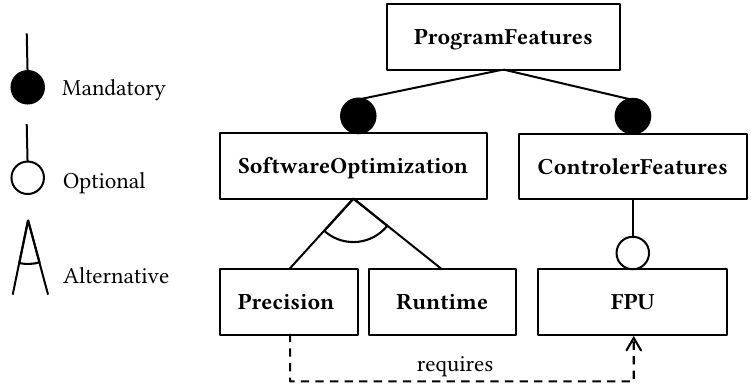}
\caption{A feature model for the product-line of \microL programs in
Figure~\ref{lst:prog3}.}
\label{fig:plFeatures}
\end{figure}

\newpage
\begin{myexampleno}
An example for a feature model is given in Figure~\ref{fig:plFeatures}. 
The product-line here offers different variants of a \microL program. 
Here, the program's variants differ, depending on whether the executing hardware platform comprises a floating point unit (feature \textit{FPU}) or not. 
Furthermore, for every variant, one needs to select an optimization (feature \textit{SoftwareOptimization}).
One of the optimization alternatives focuses on runtime efficiency (feature \textit{Runtime}), the other optimization on precision (feature \textit{Precision}). 
However, to make the high precision possible, the feature FPU needs to be selected, too (cross-tree constraint \textit{requires}).

For this feature model the set of features is:
\begin{align*}
\featuresSet &= &&\{SoftwareOptimization, ControlerFeatures, Precision,\\
& &&\ Runtime, FPU\}
\end{align*}
\end{myexampleno}

The features of the feature model shall now be used to track variability in a model product-line. 
A usual approach here, is to annotate model elements with so called \emph{presence conditions} - i.e. terms that specify to which features or feature combinations an annotated element belongs.
 
The set $\pcSet_\featuresSet$ of \emph{presence conditions} over a set of features $\featuresSet$ is defined as the set of propositional logic formulas whose atomic propositions are elements of $\featuresSet$.

\begin{definition}[Presence Condition Function]
A \emph{presence condition function} $\config:\M_M\to\pcSet_\featuresSet$ assigns presence conditions to model elements $m \in \M_M$. 
\end{definition}

\begin{lstlisting}[float,caption={A \microL product-line, annotated with the features from
Figure~\ref{fig:plFeatures}. Selecting the features \textit{FPU} and
\textit{Runtime} at the same time will result in the incorrect
variant of Figure~\ref{lst:prog2}.},label=lst:prog3] 
Program myProgramProductLine { 
	Var float myVar; 							[FPU]
	Var integer myVar;						[!FPU]
	
  myFun(p1=myVar);   	 

  Fun integer myFun(integer p1)	[!FPU | Runtime]
    { [...] }
  Fun float myFun(float p1)			[FPU & Precision]
    { [...] }}
\end{lstlisting}

\begin{myexample}
We now informally extend the example modeling language \microL with variability.
In the textual representation, features can be added to language constructs by adding a subsequent term of the syntax \textcolor{eclipseGreen}{\textit{[$<$presence condition$>$]}}.

In Figure~\ref{lst:prog3} there is a \microL program, for which there shall be different variants according to the feature model of Figure~\ref{fig:plFeatures}.
Depending on whether the executing hardware comprises a floating point unit (feature FPU), the program uses the datatype \textit{float} - otherwise only \textit{integer}. 
For runtime optimized execution (feature \textit{Runtime}) the function \textit{myFun} is declared using \textit{integer} - otherwise \textit{float} is used, in cases in which the feature FPU is available.

For the \microL program of Figure~\ref{lst:prog3}, some examples for $\config$ are:
\begin{align*}
	&\config (\text{\ttfamily\textcolor{eclipsePurple}{Var float} myVar;}) &&=
	FPU\\
	&\config (\text{\ttfamily\textcolor{eclipsePurple}{Var integer} myVar;}) &&=
	\neg FPU\\
	&\config (\text{\ttfamily\textcolor{eclipsePurple}{Fun integer}
	myFun(\ttfamily\textcolor{eclipsePurple}{integer} p1)}) &&= \neg FPU\ \vee\
	\text{Runtime}\\
	&\config(\text{\ttfamily myFun(p1=myVar);}) &&= \true
\end{align*}
\end{myexample}

\begin{definition}[Model Product-Line]
A \emph{model product-line} is a triple\\$(m, \fModel_F,\config)$, where $m$ is a model, $\fModel_F$ is a feature model with a set of features $F \subseteq \featuresSet$ and $\config$ is a presence condition function for $m$.
\end{definition}

\begin{definition}[Configuration]
\label{def:configuration}
A \emph{configuration} is as a function $k_{\fModel_{\featuresSet}}:
\featuresSet \to \boolSet$ such that $k_{\fModel_{\featuresSet}} \models \fModel_{\featuresSet}$. 
\end{definition}

This means that a configuration selects features by assigning Boolean values to each of them. 
According to this definition, $k_{\fModel_{\featuresSet}}$ is always a valid configuration, i.e.,
satisfying all constraints of the feature model.

\begin{myexample}
One possible configuration for $\fModel_{\featuresSet}$ could be:
\begin{align*}
k_{\fModel_{\featuresSet}} (f) &= 
 \begin{cases}
	 \text{true} & \text{if } f \in
	 \parbox[t]{.5\columnwidth}{\{SoftwareOptimization, ControlerFeatures, Runtime, FPU\}}\\
	 \text{false} & \text{else}\\
 \end{cases}
\end{align*}
\end{myexample}

\subsection{Symbolic Binding of Product-line Variability}
\label{sec:binding}

The introduced notions describe how to specify a model product-line by means of a core model and a presence condition function.
The objective of this paper is to simultaneously check constraints for all variants of such a model product-line -- i.e. product-line analysis.
To accomplish this, it is important to understand the effect of the presence conditions for the core model.

\begin{figure}
	\centering
        \includegraphics[width=.8\columnwidth]{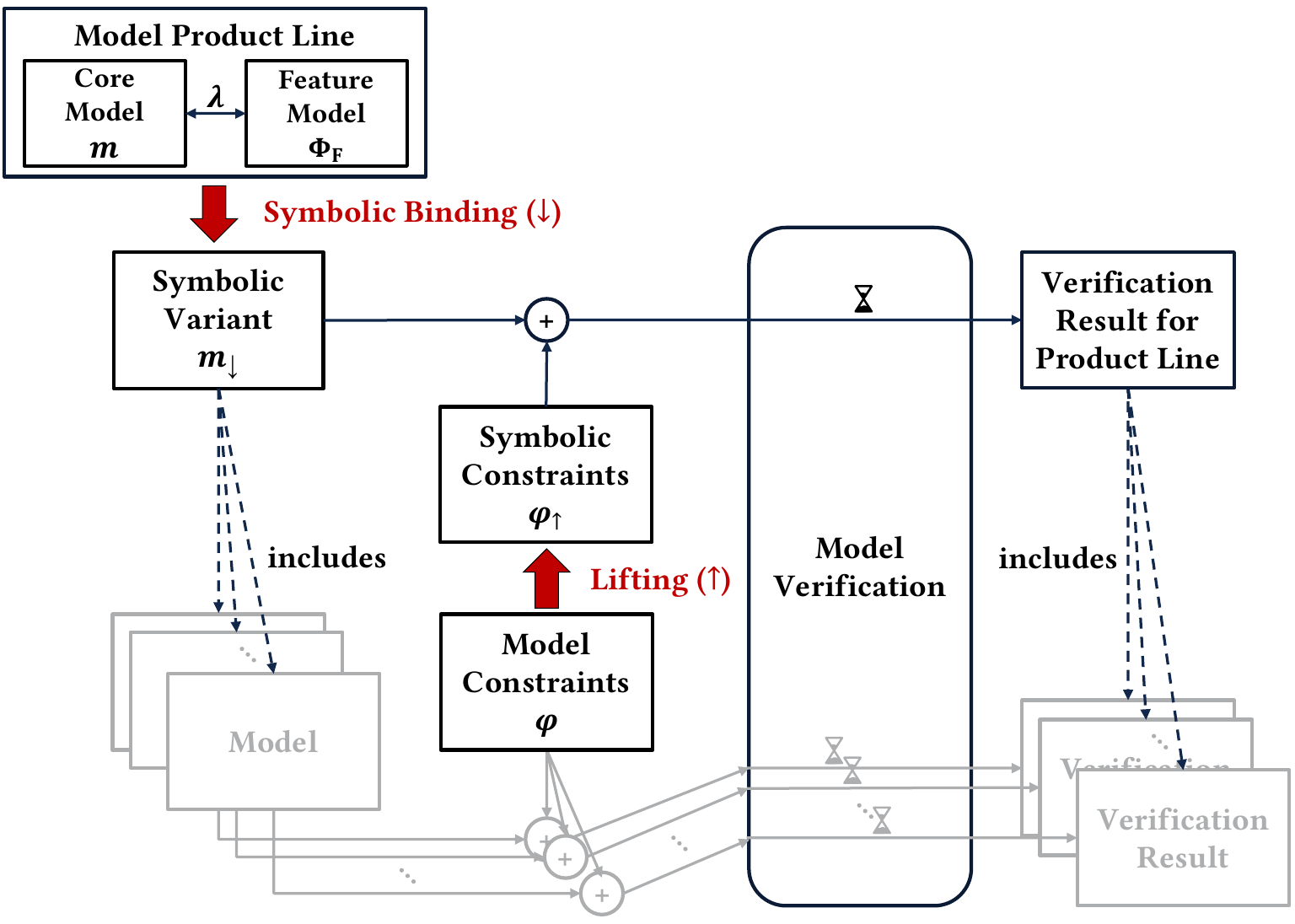}
        \caption{Refined approach for verifying a model product-line
        by symbolic binding.}
        \label{fig:lifting_binding}
\end{figure}

In this section, we will describe and formalize this effect in what we denote as \emph{symbolic binding}. 
This formalism is necessary as an auxiliary technique or pre-processing step that will be part of the lifting-based product-line analysis in this paper. 
The idea behind symbolic binding, is to encode all variability that a presence condition function $\config$ might specify for a core model $m$, immediately into one symbolic representation $m\bind$ of all possible variants.
Figure~\ref{fig:lifting_binding} illustrates this refined overview of our concept, including all of these notions.

\begin{definition}[Binding Function $\bind$]
\label{def:binding}
	Given a model product-line $(m,\fModel_F,\config)$ the \emph{symbolic binding
	function} $\bind_k$ maps the model $m$ to a model $m\bind_k$ for arbitrary
	configurations $k$ s.t. $k \vDash \fModel_F$ as follows:
\begin{align*}
m\bind_k \defeq
	\begin{cases}
		m &\text{if }m\in\mathbb B\cup\mathbb Z\cup\mathbb S\\
		\begin{cases}
		m\bind_{\mathbb{O}k} &\text{if } k\vDash\config(m)\\
		\bot &\text{else}\\
		\end{cases}
		&\text{if } m \in \mathbb{O}_M\\
	\end{cases} \\
\end{align*}
where $m\bind_{\mathbb{O}k}$ is the binding function for objects $m \in
\mathbb{O}_M$
 with $m = \{f_1, \ldots, f_n\}$ that is defined as:
\begin{align*}
m\bind_{\mathbb{O}k} \defeq \{ f_1\bind \ldots f_n\bind\}\\
\end{align*}
where
\begin{align*}
f\bind(id) \defeq \begin{cases}\begin{cases} f(id) &\text{if } k \vDash
\config (f(id))\\
\bot &\text{else}\\
\end{cases} &\text{if } f(id) \in \M_M\\
l \in \listSet(\M_M) \text{, where } \forall e
\in l.&\\ 
\hspace{4mm}e \in f(id) \wedge k \vDash \config(e)&\text{if } f(id) \in
\mathbb{L}(\M_M) \end{cases}\\
\end{align*}
\end{definition}
\vspace{-.5cm}

\begin{myexample}
The object \textit{body} $\in \M_{\mLMeta}(Body)$ of the \microL product-line of Figure~\ref{lst:prog3} has the form:

\begin{align*}
	body = 	&<varDecls \mapsto [myVarFloat, myVarInt], \\
			&funDefs \mapsto [fun1, fun2], \\
			&funCalls \mapsto [funCall]>\\
	\config(myVarFloat) =\ &FPU\\
	\config(myVarInt) =\ &\neg FPU
\end{align*}
with the symbolic binding for any configuration $k$ this would mean for the $myVarInt$ object:
\begin{align*}
	myVarInt\ \ \in\ \ body\bind_k(varDecls)\ \ \text{iff}\ \ k \vDash \neg FPU
\end{align*}
and when inserting the concrete configuration $k_{\fModel_F}$ from the example of Definition~\ref{def:configuration}, the resulting variant would be:
\begin{align*}
	k_{\fModel_F} (FPU) =\ &\true
	body\bind_{k_{\fModel_F}} =\ &<varDecls \mapsto [myVarFloat], \\
			&funDefs \mapsto [fun2], \\
			&funCalls \mapsto [funCall]>
\end{align*} 
\end{myexample}

Intuitively speaking, this binding inserts \textit{if-then-else} constructs as a distinction of cases for the existence of referred objects.
An association to an optional object will refer to this object only \textit{if} its presence condition is true - and to $\bot$, \textit{else}. 
Analogously for list associations: an object will be in a variants list \text{iff} the objects presence condition evaluates to true.

Note that throughout this definition, the configuration $k$ remains a free parameter. 
This is why we denote these functions as \textsl{symbolic} binding functions.
They could be used to actually generate a variant by instantiating $k$ with a concrete configuration. 
In the context of this work, the intention is to let a solver reason over all configurations (Section~\ref{sec:analysis} will show an exemplary SMT translation). 

\subsection{Lifting Constraints to Symbolic Variant Level}
\label{sec:lifting}

The introduction of variability into a model entails that the resulting model product-line contains model elements for \emph{several} variants at the same time.
Since the constraints for the modeling language however specify correctness for \emph{individual} models, they usually do not apply for model product-lines anymore. 

Our solution to re-enable constraint verification on model product-lines is the \emph{constraint lifting}. 
It automatically performs an extension to the constraints by means of the \emph{lifting function $\lift$}:

\begin{definition}[Lifting Function $\lift$]
\label{def:lifting}
Let $(m, \fModel_F,\config)$ be a \emph{model product-line}, the \emph{lifting function} $.\lift: \cLang \to \cLang$ is inductively defined:

 \begin{align*} 
 (\forall v \in set : expr)\lift &\defeq\begin{cases}  
 \forall v \in  set :\confForm(v) \implies  expr\lift &\text{if }
 set \in \typeSet\\
 \forall v \in set : expr\lift & \text{else}\\
  \end{cases}\\
 (expr_1 \vee expr_2)\lift &\defeq  expr_1\lift\ \vee\ 
expr_2\lift\\
 ( \neg expr)\lift &\defeq \neg (expr\lift)\\
 nav\lift &\defeq nav
\end{align*}
 \end{definition} 
The result of $\phi\lift$ is again a constraint.

The intuition behind this is that a constraint only needs to hold for combinations of model elements that are selected at the same time.
Hence, the key is the first rule: 
An expression that specifies an invariant for certain model elements only needs to hold for those model elements whose presence condition evaluates to true.

For syntactic sugar, one can rewrite these rules, of course. 
For example, the rule for the $\exists$ quantor follows immediately from the $\forall$ and $\neg$ lifting rules and would be:
 \begin{align*} 
 (\exists v \in set : expr)\lift &\defeq\begin{cases}  
 \exists v \in  set :\confForm(v) \wedge  expr\lift &\text{if }
 set \in \typeSet\\
 \exists v \in set : expr\lift & \text{else}\\
  \end{cases}\\
 \end{align*}

\newcommand{\liftColor}{black}
\begin{myexample}
When lifting the constraints of the example from Definition~\ref{def:constrLang}, $\lift$ results in (changes \textcolor{\liftColor}{\underline{highlighted}}):

\noindent``All function names are unique'':
 \begin{align*}
(\forall &f_1, f_2 \in FunctionDefinition:\\
&((f_1.funName = f_2.funName) \implies (f_1 = f_2))) \lift\\
= \forall &f_1, f_2 \in FunctionDefinition:\\
&\text{\textcolor{\liftColor}{\underline{$\config_(f_1) \wedge \config_(f_2) \implies$}}} ((f_1.funName = f_2.funName) \implies
(f_1 = f_2))
\end{align*}
``All given arguments are defined'':
 \begin{align*}
 (\forall &a \in Argument: \exists v \in VariableDeclaration:\\
&a.varName = v.varName)\lift\\ 
= \forall &a \in Argument: \text{\textcolor{\liftColor}{\underline{$\config(a)
\implies$}}} \exists v \in VariableDeclaration:\\
&\text{\textcolor{\liftColor}{\underline{$\config(v)\ \wedge$}}}\ a.varName =
v.varName
\end{align*}
``All types of all variables used in all arguments of all calls, match with the
type of the respective parameters'':
\begin{align*}
(\forall &F_{call} \in FunctionCall: \forall a \in F_{call}.args:\\
\exists &F_{def} \in FunctionDefinition: \forall p \in
F_{def}.params:\\
\exists &v \in VariableDeclaration:\\
&a.paramName = p.paramName\ \wedge \\
&a.varName = v.varName \implies \\
&v.varType = p.paramType)\lift\\
= \forall &F_{call} \in FunctionCall:
\text{\textcolor{\liftColor}{\underline{$\config(F_{call})\implies$}}} \forall a \in F_{call}.args:\\
\exists &F_{def} \in FunctionDefinition: \text{\textcolor{\liftColor}{\underline{$\config(F_{def})\
\wedge$}}}\  \forall p \in
F_{def}.params:\\
\exists &v \in VariableDeclaration: \\
&\text{\textcolor{\liftColor}{\underline{$\config(v)\
\wedge$}}}\ (a.paramName = p.paramName\ \wedge \\
&a.varName = v.varName \implies \\
&v.varType = p.paramType)
\end{align*}
\end{myexample}

Note that in Definition~\ref{def:lifting} e.g. the last rule ``$nav\lift \defeq nav$'' for navigation expressions can only be that simple, since these constraints are applied on models by means of the symbolic binding. 
This means loosely speaking that navigation expressions can only reach model elements that are \textit{present}, since the binding already made references and attributes symbolically dependent from their presence condition. 
This also applies to the second case of the quantifier rule, where $set$ is not a type, but some navigation expression.

%% file: text/smt-impl.tex
\label{sec:analysis}

The presented formalization of variability in models with the symbolic binding and constraint lifting are our basic techniques for product line analysis.
To implement it, there are different possible techniques, such as (interactive) theorem proving or SAT solving. 
In this section we show, how to utilize a SMT solver for fully automated, generic product line analysis.

The core of our technique is the formal representation of models and variability in models, that we presented in sections~\ref{sec:dsml} and~\ref{sec:variability}. 
By following the principles of this formalism, we can express models, constraints, and variability as a SMT formula and check them using a SMT solver.

The SMT formula $\zeta$ we use is of the following form:

\begin{align*}
\zeta \defeq \fModel_F \wedge \zeta_{m_M\bind} \wedge (\zeta_{\constr_1\lift}
\vee \ldots \vee \zeta_{\constr_n\lift})\\
\end{align*} 
where $\fModel_F$ is the propositional logic formula for the feature model, $\zeta_{m_M\bind}$ encodes the model $m \in \M_M$ and $\zeta_{\constr_i\lift}$ are the constrains to be checked.

Hereby $\zeta_{m_M\bind}$ represents the model $m_M\bind$ according to the notion of symbolic binding from Section~\ref{sec:binding}.
The formulae $\zeta_{\constr_i\lift}$ are the SMT representation of the lifted constraints $\constr_i\lift$ according to Section~\ref{sec:lifting} and is defined as:

\begin{align*}
\zeta_{\constr_i\lift} \defeq \neg \constr_i\lift\\
\end{align*} 

\begin{figure}
 \centering
 \includegraphics[width=.75\columnwidth]{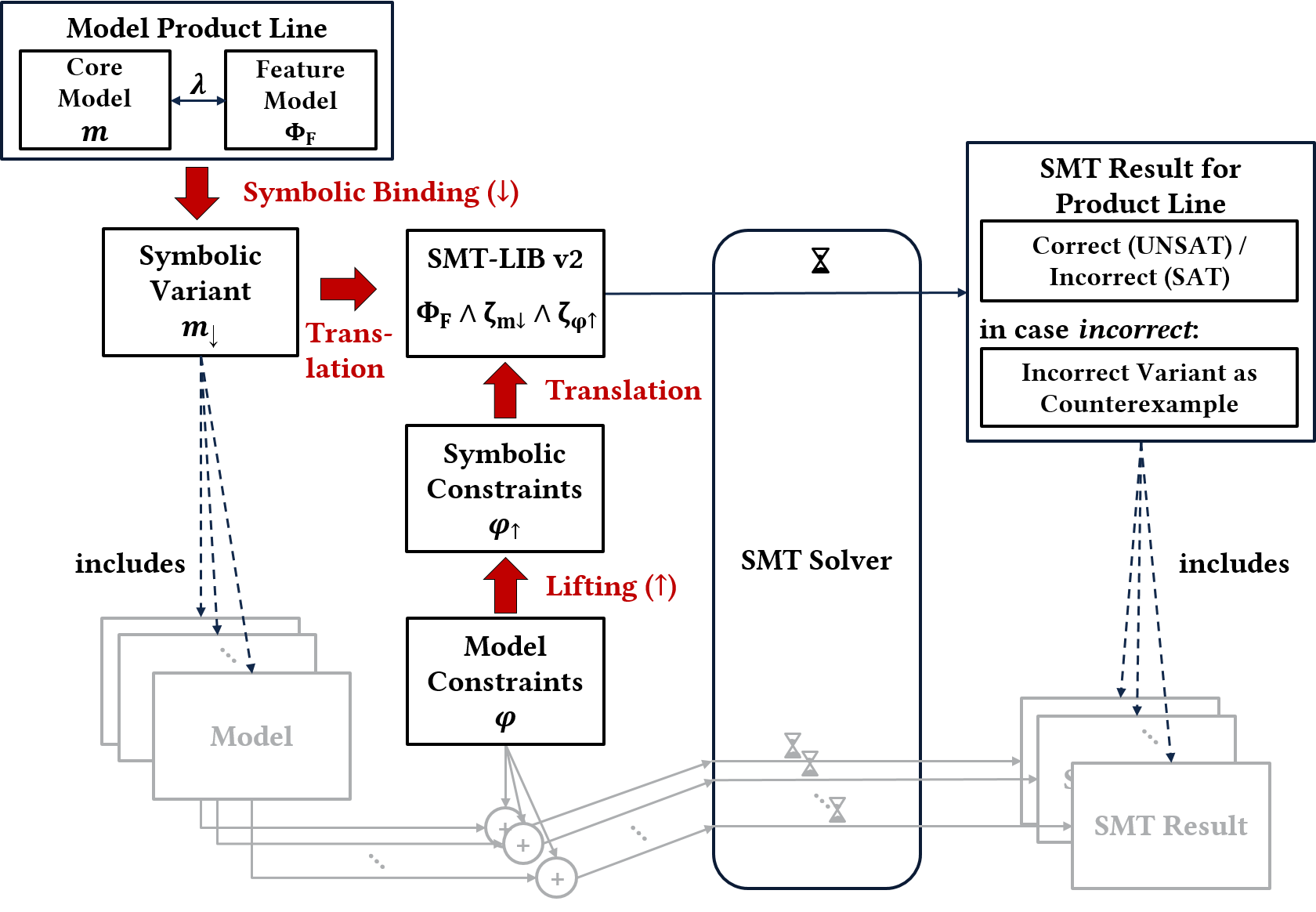}
 \caption{Principle of implementing our approach using SMT solving.}
 \label{fig:smt-impl}
\end{figure} 
 
Here, the constraints are negated in order to let the SMT solver try to find one variant that \emph{violates} a constraint.
A model product line is correct iff the solver cannot find any such variant, i.e. $\zeta \text{ UNSAT}$. 
Any model that makes $\zeta \text{ SAT}$ would fulfill at least one of the negated constraints -- with this it would be an incorrect variant. 
More precisely speaking: this model would contain a correct feature selection according to $\fModel_F$ that makes $\neg\constr_i\ \text{true}$ at the same time and with this violates $\constr_i$.

Note, that the selection of features in $\fModel_F$ is left open in $\zeta$.
With this, the SMT solvers reasons over all possible selections of features, that satisfy the feature models restrictions in $\fModel_F$.
Figure~\ref{fig:smt-impl} illustrates this implementation approach using SMT solving in our general concept figure.

SMT solvers can also provide models for such satisfiable formulae, that can be used as counterexamples, contradicting the constraints $\constr_i$.

\begin{figure}
\centering
\begin{subfigure}[b]{.58\columnwidth}
	\includegraphics[width=\columnwidth]{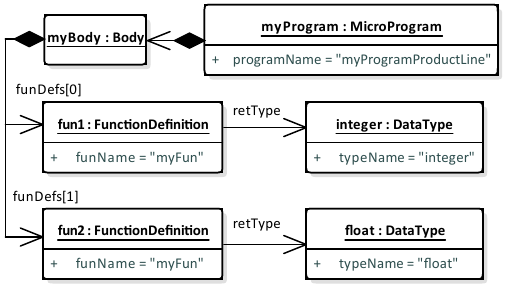}
	\caption{An excerpt of the \microL domain model that is translated for illustration.}
\label{fig:microLTranslation}
\end{subfigure}
\begin{subfigure}[b]{.4\columnwidth}
	\includegraphics[width=\columnwidth]{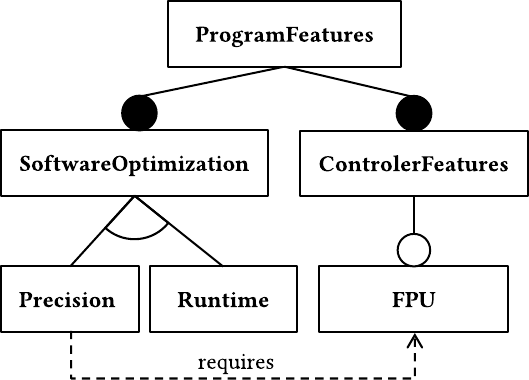}
	\vspace{.008cm}
	\caption{The feature model for the product-line.}
\label{fig:featuresTranslation}
\end{subfigure}

\caption{An excerpt of ``myProgramProductLine'' for illustrating the translation. The textual
	representation is given in Figure~\ref{lst:prog3}.}
\label{fig:translation}
\end{figure}

In the following we will present how to express $\zeta$ for a SMT solver in SMT LIB v2 synatx. 
Hereby, we use an excerpt of the running example from Figure~\ref{lst:prog3} for demonstrating the translation.
This excerpt is given in Figure~\ref{fig:translation}.

The model product-line $\zeta_{m_M\bind}$ is translated in the Sections~\ref{sec:translation_objects} and~\ref{sec:translation_variability}.
The constraint translation for $\zeta_{\constr_i}$ is given in Section~\ref{sec:translation_constraints}.

\subsection{Translation of Models}
\label{sec:translation_objects}

In Section~\ref{sec:dsml}, we formalized classes and Objects by means of set theory.
For the case of SMT an adequate representation for this are scalar enumeration datataypes for classes and their objects.
Such datatypes for the classes \textit{MicroProgram}, \textit{Body} and \textit{FunctionDefinition} from Figure~\ref{fig:microLTranslation} would be:

\begin{smt}{}
(declare-datatypes () ((MicroProgram myProgram NONE_MicroProgram)));
(declare-datatypes () ((Body myBody NONE_Body)));
(declare-datatypes () ((FunctionDefinition fun1 fun2 NONE_FunctionDefinition)));
\end{smt}

As one can see, the datatype is named according to the class names (\textit{MicroProgram}, \textit{Body}, \textit{FunctionDefinition}). 
The identifiers we chose for the scalars as \textit{myProgram} or \textit{fun1} reflect the object names from the object diagram of Figure~\ref{fig:microLTranslation}. 
This means, that for every object within a model there must be exactly one unique scalar in the datatype definition of the respective class
\footnote{An exception to this uniqueness can exist in case of inheritance. 
Here, an object might be an instance of its immediate class ans all of its super classes. 
We translated this by defining one scalar for an objects in all datatypes of all classes in the inheritance hierarchy.
In this paper we avoid this technical detail, as the general translation principal does not change.}.
Another noteworthy detail here, is the definition of a \textit{NONE} scalar for all datatypes. 
It represents the usual notion of $\bot$ or \textit{null}.

For associations and attributes, the formalization of Section~\ref{sec:dsml} used functions. 
Also in SMT functions can be used for this purpose: 

\begin{smt}{}
;; Translation of an $\color{darkgreen}\bf attribute$ in the meta model:
(declare-fun MicroProgram_programName ( MicroProgram ) String)

;; Translation of the respective instance values of the model:
(assert (= (MicroProgram_programName myProgram) "myProgramProductLine"))
(assert (= (MicroProgram_programName NONE_MicroProgram) ""))

;; Translation of an $\text{\color{darkgreen}\bf association}$ in the meta model:
(declare-fun MicroProgram_body ( MicroProgram ) Body)

;; Translation of the respective association instances from the model:
(assert (= (MicroProgram_body myProgram) myBody))
(assert (= (MicroProgram_body NONE_MicroProgram) NONE_Body))
\end{smt} 

In this example, the multiplicity of attribute and association is 1.
For higher multiplicities, we propose to use SMT sequences. 
In the diagram of Figure~\ref{fig:microLTranslation} this applies for the association ``funDefs'' of class Body:

\begin{smt}{}
;; Translation of an association $\text{\color{darkgreen}\bf list}$ in the meta model:
(declare-fun Body_funDefs ( Body ) (Seq FunctionDefinition))

;; Translation of the respective association instances in the list:
(assert (= (Body_fun myBody) (seq.++
    (seq.unit fun1)
    (seq.unit fun2))))
\end{smt} 

Besides good performances that we could see for sequences, they have the advantage that there are already theories for sequences in SMT solvers.
The Z3 solver we use, implements the theory from Bj{\o}rner et al. \cite{bjorner2012smt} and supports list length, random access to individual elements and similar.
This especially eases the translation of constraints over lists.

\subsection{Translation of Variability}
\label{sec:translation_variability}

So far, we have presented the translation of models without taking variability into account. 
This section will now introduce this variability to the SMT translation. 
The first element to be translated is the feature model. 
In Section~\ref{sec:variability} we immediately assumed propositional logic formula $\fModel_F$ for the feature model and referred to existing work. 
This means that there is no actual translation required and the feature model from the running example in Figure~\ref{fig:featuresTranslation} would be in SMT LIB v2:
  
  \begin{smt}{}
;; Declaration of Features:
(declare-const ProgramFeatures bool)
(declare-const SoftwareOptimization bool)
(declare-const ControlerFeatures bool)
(declare-const Precision bool)
(declare-const Runtime bool)
(declare-const FPU bool)

;; Mandadory Features:
(assert ProgramFeatures)
(assert (=> ProgramFeatures SoftwareOptimization))
(assert (=> ProgramFeatures ControlerFeatures))

;; Feature Hierarchy:
(assert (=> SoftwareOptimization ProgramFeatures))
(assert (=> Precision SoftwareOptimization))
(assert (=> Runtime SoftwareOptimization))
(assert (=> FPU ControlerFeatures))

;; Alternative Feature Exclusion:
(assert (=> Runtime (not Precision)))
(assert (=> Precision (not Runtime)))

;; ``Requires'' Relation:
(assert (=> Precision FPU))
\end{smt}    

As one can see, Features are translated as uninterpreted bool constants. 
The feature tree structure with the cross-tree constraint \textit{requires} are expressed by the implications in lines 10 to 25.

Based on these features, the presence condition function $\config$ can be translated straight forward as a function \textit{selected\_\textless~ClassName~\textgreater}.
For the \\ $FunctionDefinition$ of the running example this selection function would be:
\begin{smt}{}
;; Declaration of selection function analogly to $\color{darkgreen}\config$:
(declare-fun selected_FunctionDefinition ( FunctionDefinition ) Bool)

;; Translation of presence conditions for all FunctionDefinitions: 
;;    $\color{darkgreen}\config (fun1) = \neg FPU \vee Runtime$
(assert (= (selected_FunctionDefinition fun1) 
	(or (not FPU) Runtime)))
;;    $\color{darkgreen}\config (fun2) = FPU \wedge Precission$
(assert (= (selected_FunctionDefinition fun2) 
	(and FPU Precision)))
;;    $\color{darkgreen}\config (\none) = true$
(assert (= (selected_FunctionDefinition NONE_FunctionDefinition) 
	true))
\end{smt}

Note, that the \textit{NONE} element is always selected as null cannot be optional. And so is every object for which there is no presence condition defined.
In both cases, the selection function maps to $true$. 

This selection function is also relevant for the \emph{symbolic binding}. 
Symbolic binding can be formulated according to its Definition~\ref{def:binding} by using the SMT construct \textit{ite} (``if-then-else''). 
For the running example, the variability of FunctionDefinitions would result in the following translation of variability in the association from class $Body$ to class $FunctionDefinition$: 
\begin{smt}{}
;; Declaration of Association:
(declare-fun Body_funDefinitions (Body) (Seq FunctionDefinition))

;; Associtaion to existing FunctionDefinitions: 
(assert (= (Body_funDefinitions myBody) (seq.++
  (ite (selected_FunctionDefinition fun1) 
  		(seq.unit fun1) 
  		(as seq.empty (Seq FunctionDefinition)))
  (ite (selected_FunctionDefinition fun2) 
  		(seq.unit fun2) 
  		(as seq.empty (Seq FunctionDefinition)))
)))
\end{smt}

Here, list members are translated either as one-element-sequences or empty sequences - depending on their selection function (i.e. their presence condition).

\subsection{Translation of Constraints and Generation of Counterexamples}
\label{sec:translation_constraints}

Correctness constraints are already expressed in first order logic. 
We only need to negate them in order to let the SMT solver try to find one variant among all possible feature configurations that \emph{violates} the constraint.
Without this negation, the solver would search for \emph{one} feature configuration that fulfills all constraints instead of checking \emph{all} variants for one violation.

Recall the lifted constraint \textit{``All function names are unique''} from Section~\ref{sec:lifting}:   
\begin{align*}
\forall &f_1, f_2 \in FunctionDefinition:\\
&\config_(f_1) \wedge \config_(f_2) \implies ((f_1.funName = f_2.funName) \implies
(f_1 = f_2))
\end{align*}


It would be negated and translated as:

\begin{smt}{}
(assert (exists ((f1 FunctionDefinition)(f2 FunctionDefinition)) 
	(and
		;; Check selection according to lifting: 
		(selected_FunctionDefinition f1) 
		(selected_FunctionDefinition f2)
		;; In the SMT translation the NONE elements need to be excluded 
		;; from quantification:  
		(not (= f1 NONE_FunctionDefinition)) 
		(not (= f2 NONE_FunctionDefinition))
		;; The negated quantified expression from the original constraint
		;; $\color{darkgreen}(f_1.funName = f_2.funName) \implies(f_1 = f_2)$: 
		(= 
			(FunctionDefinition_funName f1)
			(FunctionDefinition_funName f2)
		(not (= f1 f2)) )
	)
))
\end{smt}

Note, that due to the implementation of null/$\none$ in SMT, there are additional clauses like line 8 ($f1 \neq \text{NONE\_FunctionDefinition}$) necessary to exclude the NONE elements from quantification.

%% file: text/sfit.tex
\label{sec:sfit}

In the previous sections, we introduced the lifting approach and applied it to the \microL language, as a running example.
In this case study, we apply it to the \sfit modeling language for production planning, that we described in previous publications~\ifthenelse{\boolean{reviewversion}}{\cite{bayha2016factoryBlind}}{\cite{bayha2016factory,kondeva2015sfit}}.

To this end we first describe the use case for which \sfit was designed in cooperation with the engine manufacturing department of \bmw, in Section~\ref{sec:manufacturing}. 
After that, Section~\ref{sec:sfit_language} presents the \sfit language and applies the constraint lifting to it.

\subsection{Variability as a Challenge in Manufacturing Planning}
\input{text/sfit-use-case}

\subsection{The \sfit Modeling Language}
\label{sec:sfit_language}

In order to address the mentioned challenges in manufacturing planning we developed the \sfit\footnote{https://www.fortiss.org/en/publications/software/sfit} modelling language in our previous work~\cite{bayha2016factory,kondeva2015sfit}.
A \sfit model captures all information about the assembly line and the product that is necessary to automatically perform the check for producibility.
As the products -- the engines in case of \bmw~-- are variable, also the \sfit model contains variability and hence is a model product-line.
In turn, every variant of this model product-line is the manufacturing planning model for one engine variant.

\begin{figure}
 \centering
 \includegraphics[width=.7\columnwidth]{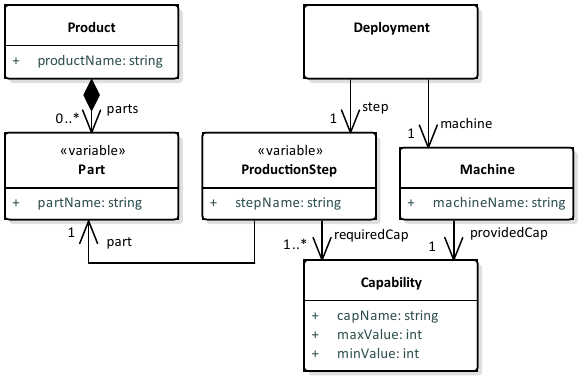}
 \caption{A simplified metamodel of the \sfit language. The original metamodel consists of about 40 classes. 
 Yet, the essence of the modelling approach is still contained.}
 \label{fig:sfit_meta}
\end{figure}

With a \sfit model being a model product-line, the check for producibility of all engine variants is an instance of product-line analysis.
The constraints for this analysis are the ones for producibility, mentioned in the previous section~\ref{sec:manufacturing}.   

For the presentation of the metamodel and constraints in this paper, we use a simplified version of the metamodel that can be seen in Figure~\ref{fig:sfit_meta}. 
The complete modeling language is larger and defined by a metamodel of about 40 classes. 
But already the simplified version here is representative, as it contains the essence of how we modeled the major use cases of the assembly planning department.

In this simplified metamodel, the product is represented by the classes \emph{Product} and \emph{Part}. 
The production process is modeled by the \emph{ProductionStep}s and the assembly lines machines correspond to the respective \emph{Machine} class.
Both -- ProductionSteps and Machines -- have an association to the \emph{Capability} class to describe manufacturing capabilities. Such Capabilities are \textit{required} in the ProductionStep and
\textit{provided} by the Machine.
Finally, a \emph{Deployment} class describes the mapping of ProductionSteps to Machines.

\begin{figure}

\begin{subfigure}{\columnwidth}
\centering
\includegraphics[width=.7\columnwidth]{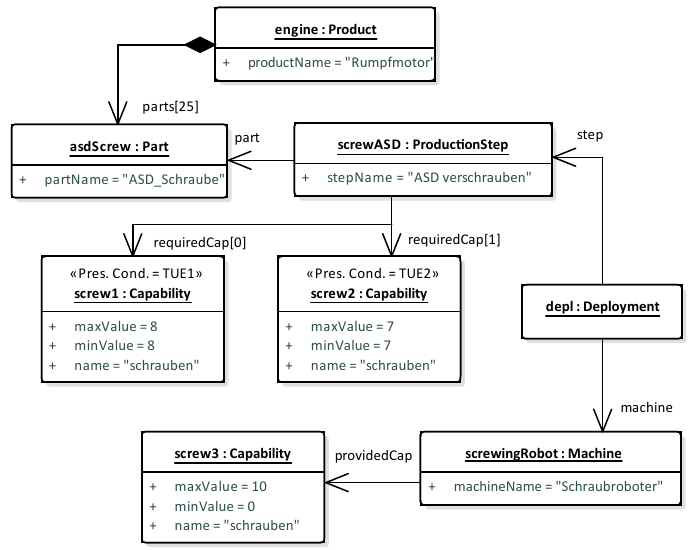}
\caption{An object diagram with a part of the engine assembly model. 
The presence conditions for optional objects are documented in the stereotype field. 
Note, that the differences \wrt the screenshots below are due to the mentioned metamodel simplification for the presentation in this paper.}
\label{fig:sfit_bmw-model}
\end{subfigure}
  
 \begin{subfigure}{\columnwidth}
\centering
\includegraphics[width=.9\columnwidth]{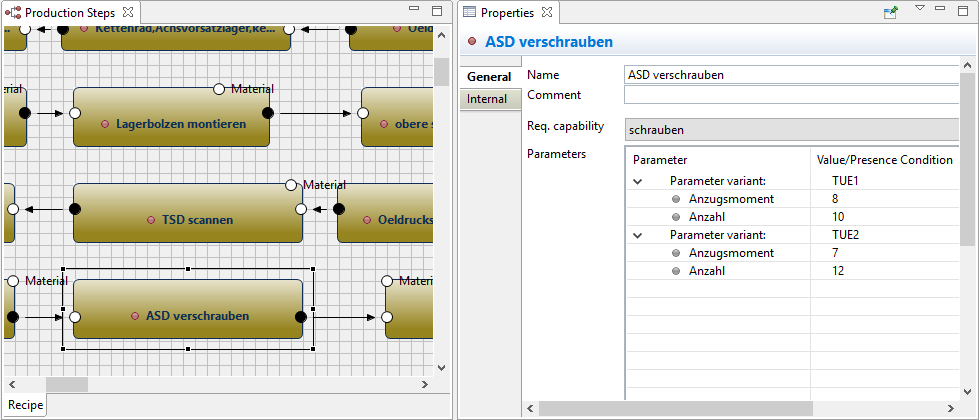}
 \caption{Screenshot from the \sfit tool. 
 The ProductionSteps require Capabilities with parameters. 
 The selected step contains variability in form of two alternative parameter sets. 
 Presence conditions -- here the features \textit{TUE1} and \textit{TUE2} -- are specified for each parameter set.}
 \label{fig:sfit_recipe}
 \end{subfigure}
 
 \begin{subfigure}{\columnwidth}
\centering
\includegraphics[width=.9\columnwidth]{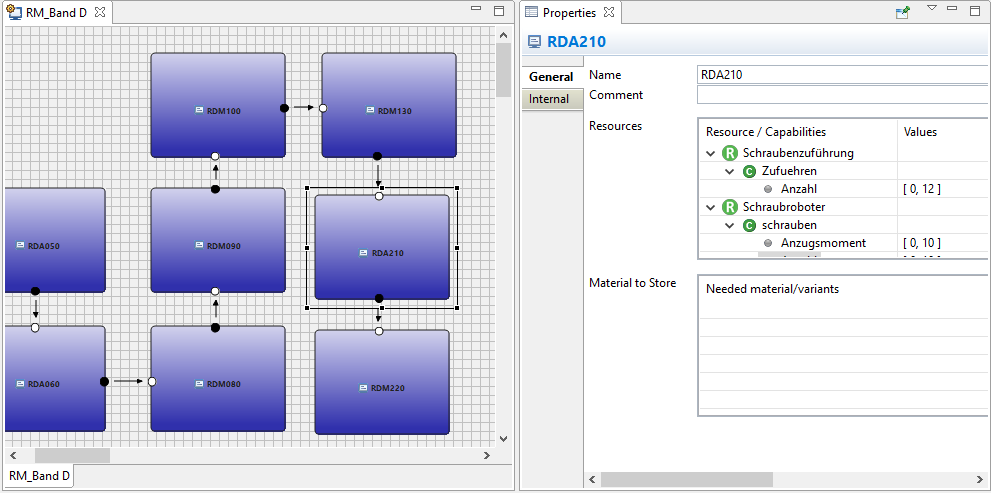}
 \caption{The part of the \sfit model for the assembly line of Figure~\ref{fig:sfit_cad-example}. 
 The blue stations contain the \textit{Machines} (``Resources'') that provide manufacturing \textit{Capabilities}.}
 \label{fig:sfit_line}
 \end{subfigure}

\caption{Excerpts of the engine assembly model.}
\label{fig:sfit}

\end{figure}

For the \sfit language we only allow variability for certain classes of model elements.
We also keep this explicit declaration of variability in the simplified metamdodel.
Here, the objects of \textit{Part} and \textit{ProductStep} may be optional -- indicated by the stereotype \textit{<<variable>>}.
\andreas{Remove before flight: Is the readability aspect important in the end?}
In this paper this mainly serves for improved readability of the code fragments, later on. 
Besides readability such limitations of variability are out of scope of this paper, yet they also do not conflict with our approach. 

We modeled the BMW assembly line of Figure~\ref{fig:sfit_cad-example} and a simplified engine product-line in \sfit
\footnote{The simplification mainly concerns the number of features.
The complete feature model is confidential\andreas{Clearify with BMW if this formulation is ok.} and could not be used for the case study. 
Yet, for the runtime analysis in the evaluation of Section~\ref{sec:evaluation} we will also use another model with more features.}.    
Figure~\ref{fig:sfit} shows parts of this model. 
Here, Figure~\ref{fig:sfit_bmw-model} contains an excerpt of the object diagram that is an instance of the simplified metamodel.
Besides this formal example, there also are screenshots from the \sfit tool in Figures~\ref{fig:sfit_recipe} and~\ref{fig:sfit_line}.

For the producibility check -- i.e. the product-line analysis -- we formalized the correctness constraints for producibility.  
Some correctness constraints hereby are:

\noindent Constraint 1: \textit{``For all Parts of all Products, there is a ProductionStep that assembles the Part''}:
\begin{align*}
\forall &prod \in Product: \forall part \in prod.parts:\\
&\exists step \in ProductionStep: \\
&\hspace{.25cm}step.assembledPart = part
\end{align*}

\noindent Constraint 2: \textit{``All ProductionSteps are deployed to a Machine''}:
\begin{align*}
\forall &s \in ProductionStep:\\
&\exists d \in Deployment: d.step = s
\end{align*}

\noindent Constraint 3: \textit{``All ProductionSteps are deployed to a Machine, that can fulfill the ProductionSteps required Capability''}:
\begin{align*}
\forall d &\in Deployment:\\
&d.step.requiredCap.name =\\
&d.machine.providedCap.name\\
\wedge (&d.step.requiredCap.maxValue \geq\\
&d.machine.providedCap.minValue\\
\vee &d.step.requiredCap.minValue \leq\\
&d.machine.providedCap.maxValue)
\end{align*}

As an example for lifting on these constraints, the following is the result for lifting Constraint 1 \textit{``For all Parts of all Products, there is a ProductionStep that assembles the Part''}:
\begin{align*}
\forall &prod \in Product: \config(prod) \implies \forall part \in prod.parts:\\
&\config(part) \implies \exists step \in ProductionStep: \\
&\hspace{.25cm}\config(step) \wedge step.assembledPart = part
\end{align*}

When negating and translating this to SMT, the result is:

\begin{smt}{}
(exists ((prod Product)) (exists ((p Part)) 
	(and 
		(selected_part p)
		(seq.contains (Prod_parts prod)(seq.unit p)) 
		(forall ((s ProductionStep)) 
			(=> 
				(and(selected_step s)) 
				(not (= (Step_part s) p)))))))
\end{smt} 

Note, that this lifted constraint does not need to check for the selection of \textit{Product prod} as these objects cannot be variable in the \sfit implementation. 
In contrast to Parts and ProductionSteps. 

As mentioned initially, the \sfit metamodel we present above is simplified for better readability and understanding.
To give an impression of the constraints for the non-simplified metamodel, the equivalent for Constraint 3 would be:
\begin{align*}
&\forall depl \in ProcessStepStationDeployment. \\
&\forall pisStep \in depl.processStep.restrictedParameterInstanceSets.\\
&\hspace{.4cm}\exists res \in depl.station.ressources.\\ 
&\hspace{.4cm}\exists cis \in res.capabilityInstanceSets.\\
&\hspace{.4cm}\exists  pisRes \in cis.parameterInstanceSets\\
&\hspace{.8cm}depl.processStep.requiredCapability = cis.capability\ \wedge \\
&\hspace{.8cm}pisStep.parameter = pisRes.parameter\ \wedge\\ 
&\hspace{.8cm}(pisStep.valueSet.max < pisRes.valueSet.min\ \vee \\
&\hspace{.8cm}\ pisStep.valueSet.min > pisRes.valueSet.max)
\end{align*}

The product-line analysis algorithm that had previously been implemented for this constraint alone in course of our previous work~\cite{bayha2016factory} comprises 879 lines of complex Java code -- without taking shared utility methods into account. 
All of this complex code would not have been necessary to be written, tested and maintained in the first place, if the approach of the present paper would have been used.

%% file: text/sfit-use-case.tex
\label{sec:manufacturing} 

The industrial background for this case study is the assembly of combustion engines at \bmw. 
The considered product-line hereby models the assembly of inline internal combustion engines. 
Hereby, there are numerous variants of these engines which are differentiated by several features as the number of cylinders, the fuel type or the performance level. 
The engines of this product-line are designed in a modular fashion which means they share a high number of common parts. 

\begin{figure}[]
 \centering
 \includegraphics[width=\columnwidth]{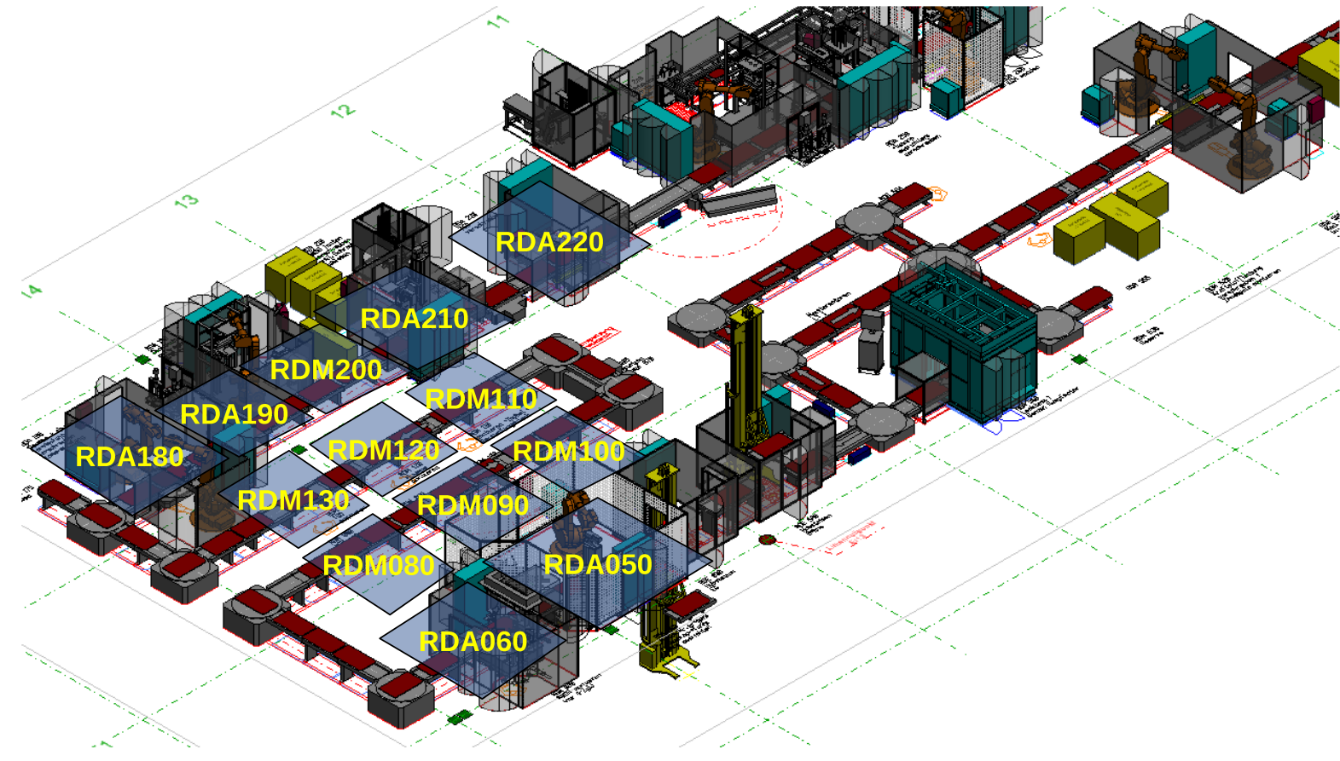}
 \caption{CAD drawing for an engine assembly line of \bmw. We modeled a part of
 this line in \sfit to perform automatized producibility checks.}
 \label{fig:sfit_cad-example}
\end{figure}

The tasks of assembling these engines is to handle, place and join single parts to a full product.  
To do so certain resources are required to execute the required assembly processes. 
An example for this could be an automated tightening robot that is used to join to parts using a screw. 
The engines are assembled on high volume assembly lines in a worldwide production network. 
For this paper we considered one of these assembly lines that can be seen in Figure~\ref{fig:sfit_cad-example}. 

As the construction of these assembly lines require major funds, their lifecycle usually exceeds multiple product lifecycles.
Thus, once an assembly line is installed it results into requirements for the producibility of all future engine variants.
All those assembly lines are designed by human assembly planners and one task for them is to assure the producibility of new variants. 
Therefore, they need to find out whether new or changed engines can be assembled on the assembly line with the resources given. 
If this is not the case the planner has to identify possible solutions to resolve the producibility conflict. 
 
This producibility check includes to check several constraints that shall be fulfilled by all variants and these constraints can appear on different levels. 
For example, on the level of the assembly line the planners have to ensure that it is capable to output the required volume of engines and the assembly sequence is not violated.
On the level of each resource typical constraints are geometric dimensions as well as for instance mass, torque, or force. 
An additional constraint can be the number of parts variants that can be stored or handled in the assembly stations as stated by M\"uller et al. in~\cite{muller2017lean}.
 
To check whether the assembly lines' resources and engine variants fulfill these producibility constraints is a significant and reoccurring task in assembly planning. 
If a constraint is violated the planner can either enforce a product change, a technical change of resources, a change of the allocation or a combination of the three as presented by M\"uller et al.~\cite{muller2015modell}.
State of the art is to manually check the fulfillment of all constraints.
The \sfit modelling tool that we will present in the next Section~\ref{sec:sfit_language} was developed to answer this produciblity question by automizing the checks for several constraints.

%% file: text/evaluation.tex
\label{sec:evaluation}

In order to give a possible technology mapping for our approach we did the SMT implementation for the EMF framework as presented in Section~\ref{sec:analysis}.
After that Section~\ref{sec:sfit} introduced the \sfit modeling language to which we applied our approach as a case study. 
Now, we use this SMT-based implementation and the \sfit case study to evaluate the applicability, correctness, and scalability of our approach.

To this end, Section~\ref{sec:sfit_models} gives an overview about the model product-lines we use for this evaluation.
After that, Section~\ref{sec:correctness} evaluates the correctness before Section~\ref{sec:scalability} demonstrates the scalability of our approach based on the SMT implementation.
Finally, Section~\ref{sec:limitations} discusses the limitations of our approach and possible threats to validity of this work.  

\subsection{Models Used for Evaluation}
\label{sec:sfit_models}

The six model product-lines we used for evaluating our approach are based on three initial model product-lines: 
one small, artificial example and two industrial assembly planning models of two different companies.
As none of these three model product-lines violated the correctness constraints, we introduced failures and came up with additional three incorrect model-product lines.
These introduced failures were (i) incorrect presence conditions to add failures resulting from variability and (ii) changed values of attributes to add purely domain specific issues.
Table~\ref{tab:models} gives an overview about all six correct and incorrect model product-lines. 

\begin{table*}[]
\centering
\begin{tabular}{|l||c|c|c|}\hline
                        		&\multicolumn{3}{c|}{Number of Elements} 										\\
Model                        	& Features (of these optional)	& Objects 				& Presence Conditions	\\\hline\hline
Pen Example (correct)      		& \multirow{ 2}{*}{3 (2)}  		& \multirow{ 2}{*}{81}	& \multirow{ 2}{*}{23}	\\\cline{1-1}
Pen Example (incorrect)    		&   							&    					&    					\\\hline\hline
Motor Body (correct)    		& \multirow{ 2}{*}{3 (2)}	  	& \multirow{ 2}{*}{562}	& \multirow{ 2}{*}{172}	\\\cline{1-1}
Motor Body (incorrect)  		& 	 							&   					&  						\\\hline\hline
Steam Cooker (correct)			& \multirow{ 2}{*}{28 (21)}		& \multirow{ 2}{*}{1227}& 103 					\\ \cline{1-1}
Steam Cooker (incorrect)		&  							  	& 						& 110 					\\ \hline
\end{tabular}
\caption{Overview about the analyzed model product-lines from the \sfit case study.}
\label{tab:models}
 \vspace{-.2cm}
\end{table*}

The artificial example was created for testing during development.
It models the production of a product-line of pens. 
As table~\ref{tab:models} shows, this small model comprises 81 model elements (i.e. objects) of which 23 are optional due to presence conditions.
These presence conditions link model elements to three features of which one is mandatory and two are optional.  

The second model product-line is a part of the BMW automotive engine assembly as introduced in the case study of Section~\ref{sec:sfit}. 
It was done by the company itself and captures most of the manufacturing planning for a product-line of engines.
In comparison to the model size, there is only little variability, as only a part of the product-variability was taken into account.
For the 562 model elements here, this means that 172 of them are variable via presence conditions. 
The feature model comprises three features of which two are optional.  

For the third model product-line we used the SFIT language to capture the manufacturing of a product-line of steam cookers from the home appliances manufacturer Miele. 
Here, we extracted the assembly planning model product-line from textual assembly instructions of Miele.
This information was grouped per assembly station and contained all assembly steps to be performed for all the variants.
The resulting model-product line consists of 1227 model elements of which 103 are variable with presence conditions.
With a feature model of 28 features in total and 21 optional features, this is our biggest model for evaluation -- both in size and variability.

\subsection{Correctness}
\label{sec:correctness}

To evaluate correctness, we used all three of the previously mentioned initial model product-lines.  
All of them are correct w.r.t. the producibility constraints. 
This is not surprising, as the respective products of the industrial models were already in production.
Also the preexisting, specific and hand-implemented analysis algorithms of the \sfit tool confirmed that there were no producibility issues in any of them.

As mentioned before, we constructed an invalid model product-line from each of these three correct ones in order to be able to also find failures.
To this end, we introduced failures into the models by changing the presence conditions or introducing additional ones.
Also attributes were changed to introduce domain specific violations of the constraints.
With this, we received three incorrect model product-lines in which some failures purely result from variability (e.g. missing dependencies) and others are real domain specific problems (e.g. invalid combinations of attribute values).

The SMT-based implementation of our approach was able to find all of these introduced failures.
For the constraints 2 and 3 we could also compare the analysis results with the results of the specific analysis algorithms that had already been implemented in the SFIT tool.
Here there were no discrepancies -- all failures were equally found by both: the new generic analysis and the old hand-implemented algorithms.
For Constraint 1 such a comparison was not possible, since there had not been any analysis before. 

For the correct three initial model product-lines there were no failures found -- neither by the old algorithms, nor by the new generic analysis.

With this, our experiments showed that the generic analysis neither produced false negative, nor false positive results.   

\subsection{Runtime Analysis and Scalability}
\label{sec:scalability}

For each model and each constraint, we measured the runtimes for the respective analyses of our SMT-based implementation.
Hereby, we executed the analysis for every constraint individually as well as for all constraints in one run.  

The resulting runtimes can be seen in Table~\ref{tab:runtime}. 
For all models and constraint set, the table gives the total runtime of the analysis (SMT translation and solver). 
The pure solvers runtimes are given in parenthesis, too.
We used the SMT solver Z3 and performed our experiment using on a standard PC, equipped with an Intel i7-6700HQ CPU with 4 cores @ 2.60GHz and 16GB RAM.

As expected, the largest model with the most optional features has the longest runtime - more than twice as long as for the medium-sized models. 
The difference between valid and invalid models does not seem to be significant. 
Also checking all constraints in one run does not significantly change the runtime w.r.t. the single constraints. 
Here, the runtime seams to correspond to the longest runtime in the case of checking the single constraints individually.      

Without putting much effort on optimization, these runtimes of the SMT-based implementation are already very acceptable. 
It is definitely possible to continuously check a model product-line during editing in order to identify failures immediately after introducing them.  

With these runtimes for the exemplary implementation, we could demonstrate the scalability of our approach also for bigger model product-lines. 

\begin{table*}[]
\centering
\begin{tabular}{|l||l|l|l|l|}\hline
                        		&	\multicolumn{4}{c|}{ Runtime (SMT Solver alone in braces) [sec.]} \\
Model                        	& Constraint 1 	& Constraint 2 	& Constraint 3 	& All Constraints \\\hline\hline
Pen Example (valid)      		&   0.13 (0.05) 	& 0.10 (0.04)	& 0.14 (0.06)	& 0.14 (0.07) \\\hline
Pen Example (invalid)    		&   0.16 (0.07) 	& 0.12 (0.05)	& 0.17 (0.08)	& 0.15 (0.08) \\\hline\hline
Motor Body (valid)    			&  0.70 (0.27)	& 0.60 (0.21)	& 1.50 (0.80)	& 1.46 (1.05) \\\hline
Motor Body (invalid)   			&  0.95 (0.46)	& 0.71 (0.34)	& 1.56 (1.03)	& 1.02 (0.64) \\\hline\hline
Steam Cooker (valid)			&  3.79 (2.72)	& 1.85 (1.03)	& 3.26 (2.09)	& 3.65 (2.75) \\ \hline
Steam Cooker (invalid)			&   4.23 (3.03)	& 2.24 (1.39)	& 3.90 (2.81)	& 4.01 (3.02) \\ \hline
\end{tabular}
\caption{The runtimes for analyzing the constraints. The first three columns show individual solver runs for every constraint. The last column gives runtimes for an execution checking all constraints in one solver run.}
\label{tab:runtime}
 \vspace{-.2cm}
\end{table*}

\subsection{Limitations and Threads to Validity}
\label{sec:limitations}

Our approach aims to describe the analysis problem as universally as possible. 
With this we want to enable the applicability to different technologies, tools and languages.
A limiting factor hereby could be the expressiveness of first order logic (FOL) that we use for the representation of constraints.
If the used constraint language should be more expressive as FOL, the approach would probably have to be extended accordingly.
We do not expect this to be an issue in practice, as also the widely used Object Constraint Language (OCL) can be translated to first order logic as can be seen in the work of Beckert et al.~\cite{beckert2002translating} or Kuhlmann and Gogolla in~\cite{kuhlmann2012uml}.

Similarly, the modelling language needs to be able to be mapped to our formalization.
Also here, we do not expect this could be an actual problem, as the formalized concepts of objects, attributes and association are absolute standard concepts in modelling.

More noteworthy is the fact, that of course the constraints to be lifted need to be correct in the first place to yield a correct analysis.
This might be an obvious and trivial insight. 
Yet, it is worth mentioning that the formulation of correct constraints should be reserved to experts, that deeply understand a respective metamodel.     

An important aspect we want to underline here is that our notion of analysis aims at analyzing static models. 
This means, that static aspects as for example type correctness could be verified, but not dynamic aspects as e.g. an variables' possible value range during execution.
For such kind of analyses, we refer to related work in Section~\ref{sec:related_work}.

A thread to validity \wrt this work is, that we evaluate our approach with one verification technique and one modelling language -- namely SMT solving and the SFIT language.
We cannot guarantee, that the scalability would be similar for other implementations as ILP solving or theorem proving.
Also other modelling languages and constraints could be different in the runtime behavior in case they should be less well suited for SMT solving.
We are about to further investigate this applicability in a second modelling tool for systems engineering in our future work.

%% file: text/related-work.tex
\label{sec:related_work}

A very good overview on existing literature on product-line analysis was done by Thum~et~al.~\cite{thum2014classification}. 
This work not only gives an overview of the field, but also comes up with a classification of product-line analysis strategies. 
The three major categories hereby are \textit{Product-based Analyses}, \textit{Feature-based Analyses} and \textit{Family-based Analyses}.
Very similar categories are also defined by Apel~et~al. in~\cite{apel2013strategies}. 
Our work clearly belongs to the family-based strategies, as our models are family artifacts that implement many features in one module and since we also analyze for all features simultaneously.
In the remainder, we concentrate on family-based approaches accordingly.

Another noteworthy literature survey was recently published by Pol'la~et.~al~\cite{pol2020analysis}. 
This study is broader and also takes into account forter kinds of analyses as e.g. the analysis of features models alone.
As one result of their work, the authors identify eight challenges for future research of which we contribute to Challenge 1 (Formal Framework), Challenge 3 (Benchmarks) and Challenge 7 (Support for Domain as well as Application Engineering).     

\subsection*{Language independent product-line analysis}
There is existing work on \emph{language independent} product-line analysis.
Kastner~et~al. propose a generic product-line syntax check for arbitrary textual languages in~\cite{kastner2009guaranteeing}. 
The analysis implementation can automatically be adapted to new languages by providing an annotated grammar that defines the syntax.
The analysis then is able to verify for a domain artifact, that all variants that can be generated form it, are syntactically correct.
In contrast, our approach does not focus on textual syntax, but the static structure of a model.
Apel~et~al.~\cite{apel2010language} abstracted from work on product-line type checking and give an algorithm for language independent reference-checking for product-lines.  
We in contrast are not limited to one kind of constraint or analysis method.
The dissertation~\cite{mazo2011generic} talks about a generic approach to verify product-line models. 
Here, the modeling language is generic, the constraints however are predefined and checked by an individual algorithm each.
Similarly, Lopez-Herrejon and Egyed give constraints for safe composition in multi-view models with variability in Multi-View Models~\cite{lopez2010detecting}. 
The authors also implemented an analysis to check these constraints. Yet, the implementation was manually done per constraint. 
New constraints could not be checked automatically.  
Also Bilic~et~al. use fixed constraints in~\cite{bilic2020detecting}. 
The authors identified these constraints for the SysML and their analysis can not only detect possibly incorrect model variants but also ambiguities in variants.

We are more generic than these above contributions, since for our approach the constraints are generic and can be defined for each language individually.

The work of Famelis~et~al.~\cite{famelis2017software} does not only consider a fixed set of constraints.
Yet, all constraints need to be defined immediately in the level of product-lines.
I.e. one needs to specify what correctness means for a product-line immediately, whereas our approach is based on a notion of correctness for product variants. 
The authors furthermore distinguish four categories of properties that also take design uncertainty into account. 
I.e. they allow constraints that \emph{might} hold only for sets of variants.
The constraints that we target belong to the category ``Necessary for all products of a product-line''.

Also the work of Barner~et~al.~\cite{Barner2016} takes technical design uncertainties in product-lines into account. 
In their work, constraints are verified during the synthesis of correct variants within the design space that results from uncertainty.  
We in contrast consider managed variability within a product-line, while their work is based on a different notion of variability and does not target the analysis of predefined model product-lines. 

The paper~\cite{heidenreich2009towards} of Heidenreich~et~al. is closer related to our lifting approach. 
It proposes to use a constraint language for EMF models with the aim to check domain artifacts against these constraints.
However, this work remains a proposal - to our knowledge there is no publication with a language definition or implementation.

\subsection*{Analysis of product-lines of languages}
A dedicated field of research is the analysis of families of related modelling languages -- i.e. language product-lines, one meta level above our work.

Gr\"onniger and Rumpe define the constituents of such language product-lines including concrete and abstract syntax as well as semantics and their variability in~\cite{groeninger2011modeling}.
The focus of their work is on variability in a modelling languages semantics. 
To this end, the authors define a concept of \textit{semantic language refinement} and how to prove the correctness of such refinements in variants.
In contrast to these authors work, we do not allow variability in formal semantics of modelling languages but focus on structural variability in model instances. 

In a similar direction, the work of Maoz et al. in~\cite{maoz2011semantically} presents an analysis for semantic consistency of object diagrams \wrt their class diagrams.
The authors present a feature model that specifies the possible variability in the semantics of class diagrams using 32 features.  
The analysis can then be configured along these features.
In a certain way this is similarly to our work, as these authors present a way to use an analysis for different modelling languages.
However, in contrast to our work (i) the focus is on variability in semantics and (ii) the variability is fixed an encoded in the analysis itself.   

Guerra~et~al. use a technique, that is similar to our symbolic binding in ~\cite{guerra2020property}. 
So-called feature-explicit metamodels (FEMM) are generated from 150\% metamodels and feature models in order to analyze the instanciability of the metamodel product-line. 
However, in their 150\% metamodels the \textit{properties} to be checked themself are part of the product-line and already get assigned presence conditions. 

An overview of further literature on product-lines of languages can be found in ~\cite{mendez2016leveraging}~\cite{cengarle2009variability}.

\subsection*{Model checking and SAT solving for product-line analysis}
Another related direction is applying \emph{model checking} for product-lines, as in the work of Classen~et~al.~\cite{classen2009model} or Lauenroth~et~al.~\cite{lauenroth2009model}.
While in contrast to us, this area focuses on system states and behavior, it also reasons over a whole family of systems.
The paper~\cite{ben2015symbolic} of Ben-David~et~al. also researches this field. 
Noteworthy w.r.t. our paper is that they also use SAT-based approaches and think about modifying the verified properties instead of the model in their future work.
In~\cite{apel2011detection}, Apel et al. present a similar approach for the detection of feature interactions, using model checking. 
However, the authors use feature modules for the definition of their product-lines. 
Therefore, the modeling approach is different from ours -- we do not assume the product-line or the specification to be structured in feature modules but target 150\% product-line engineering.

The publication~\cite{young2020variational} of Young~et~al. shows a way how SAT solvers can be used to incrementally check formulas with variability.
Their work is complementary to our solution as it could improve solver runtimes when analyzing a changed model product-line a second time.
We are confident, that a similar technique could also be integrated in our SMT based implementation to improve runtimes of incremental analysis runs.
We consider to target this in course of the optimizations in our future work. 

\subsection*{Lifting of product-line analysis}
There are some works, that also propose ways to reuse existing analysis methods by some \emph{lifting}. 
Post~et~al. propose this lifting in~\cite{post2008configuration} for the domain artifacts themselves. 
Similar to what we call \emph{symbolic binding}, C code artifacts are extended in such a way, that they have the configuration information encoded using native C language constructs. 
This enables using a standard C model checker - CBMC in their case.
However, this analysis is only applicable for C code domain artifacts. 

In contrast to lifting the artifacts, Mitgaard~et~al.~\cite{midtgaard2014systematic} show a way to lift verification methods themselves to product-line level. 
The approach of their work is to lift the derivation of abstract interpretations to product-line level.
Bodden~et~al. lift static analysis of source code such that it can be reused for analyzing software product-lines~\cite{bodden2013spllift}. 
Hereby they use the IDE solver \textit{Heros} and apply their solution to Java-based software product-lines.   
Both works are complementary to our work, since they aim on data flow analysis, whereas we are interested in static properties of models.

Also the work of Alwidian and Amyot in~\cite{alwidian2020union} lifts an existing analysis technique itself. 
The authors enhance an existing method for the analysis of individual goal models to also support the simultaneous analysis of families of goal models.
In the particular case of their work these families are represented as so-called union models which capture variability and versioning of goal models. 
Our approach does not support versioning. 
Instead, we are not limited to one kind of models or modelling language and support the usual notion of annotation-based 150\% variability.

Another kind of lifting is presented by Salay~et~al.~\cite{salay2014lifting}. 
This work is about lifting model transformations instead of constraints. 
Yet it is also interesting to be mentioned here, since the authors' notion of lifting is similar to ours:
after a lifted model transformation is applied, the resulting product-line will yield the same variants, as if the original transformation would have been applied to those.    
 
Lifting is also foreseen in the formal framework that Castro~et~al. present in their work~\cite{castro2021formal}.
This framework aims on describing different kinds of product-line analyses with their properties each.
The authors however conceived their framework for \textit{software} product-lines and properties of software and transition systems. 
The relation to \textit{model} product-lines is out of scope of this work and remains open.

\subsection*{OCL based product-line analysis}

Buchmann~et~al. define correctness constraints in OCL for the correctness of UML models in their publications~\cite{buchmann2012ensuring} and~\cite{buchmann2014mapping}.
Similar to other mentioned publications above, all constraints are fixed in their work and target notions of correctness that originate from the UML itself.
Interesting in comparison to our approach is, that all those constraints are defined immediately on product-line level.

The closest work we know was done by Czarnecki~et~al. in~\cite{czarnecki2006verifying}.
Here OCL invariants are used for the specification of domain specific correctness constraints.
Instead of being lifted, the semantics of the OCL constraint is redefined there.
The result of interpreting a constraint according to this redefined semantics is not a single value per constraint, but all possible values with presence conditions each. 
Also the template interpretation step requires to explicitly evaluate all variation points which are relevant for the constraint to be checked in order to come up with all possible values.
This might be a downside in practice, as it does not make use of the strong heuristics that are implemented in state-of-the-art SMT solvers.   
A main difference is that the authors analysis is fixed and limited to the capabilities of OCL and the template interpretation of OCL.
We are language independent and also target e.g. the upcoming SysML v2 language, which is not based on UML or OCL.

%% file: text/summary.tex
\label{sec:summary}

We presented a language independent approach to analyze model product-lines for correctness w.r.t. language specific constraints.
To this end, we gave a general formalization of the product-line analysis problem for structural invariants of models and a generic solution to perform such analyses. 
Our major contribution hereby is a way to prepare the constraints for variants by a \emph{lifting function}, such that they are applicable to a model product-line.
With this, the lifted constraint can be verified on the model product-line to simultaneously check the correctness of all variants that can be generated. 
As an auxiliary technique for the constraint lifting and the analysis, we introduced \emph{symbolic binding}.
Hereby, all variability is encoded in such a way that it can be translated to the used verification mechanism.

We presented how to use this approach for implementing generic product-line analysis using SMT solving.
This exemplary implementation is based on the Z3 Java API and automatically translates EMF model product-lines to SMT. 
Finally, a case study illustrates the application to the \sfit modeling language from the domain of manufacturing planning.
The analysis of two industrial \sfit model product-lines based on production planning data by the BMW Group and Miele demonstrates the scalability of our approach.

Our approach is not only independent from a specific metamodel, but also does not depend on the language or the theorie(s) that are used in the constraints.
It can even be applied with different underlying verification mechanisms as SMT solving or theorem proving.
With this it cannot only be used with UML/OCL but also applied to other modelling approaches.

A particularly interesting field of application will also be the upcoming SysML v2 standard, which will not be based on UML but the KerML language.
Particularly, as for SysML/KerML there is no constraint language as OCL defined, yet.
Also in future work the \afocus tool~\ifthenelse{\boolean{reviewversion}}{\cite{aravantinos2015autofocusBlind}}{\cite{aravantinos2015autofocus}} and modeling language shall be enhanced with variability and our product-line analysis. 
This will demonstrate the applicability to another modeling language and also enable another runtime analysis for the large model product-lines we did in \afocus.

Concerning our exemplary implementation using SMT solving we did not apply any optimizations, yet.
We are confident that the runtimes could even be improved by optimizing the SMT input or incorporating iterative techniques. 